\documentclass[12pt, fleqn]{article}
\usepackage[cp1251]{inputenc}
\usepackage{latexsym,amsfonts,amssymb}
\usepackage{graphicx}

\usepackage{amsbsy}
\usepackage{amsmath}
\usepackage{epsf}
\usepackage{cite}
\newtheorem{theo}{Theorem}
\newtheorem{remark}{Remark}

\newcommand{\bt}{\begin{theo}}
\newcommand{\et}{\end{theo}}
\newcommand{\bd}{\begin{displaymath}}
\newcommand{\ed}{\end{displaymath}}

\newcommand{\lf}{\left}
\newcommand{\rg}{\right}

\newcommand{\be} {\begin{equation}}
\newcommand{\ee} {\end{equation}}
\newcommand{\ba}{\begin{array}{l}}
\newcommand{\ea} {\end{array}}

\newcommand{\p} {\partial}

\newcommand{\al} {\alpha}
\newcommand{\lbd} {\lambda}

\begin{document}

\begin{center}
 {\Large \bf Lie symmetries,  reduction  and exact solutions \\
 of the (1+2)-dimensional nonlinear problem \\
 modeling the solid  tumour growth }
\medskip

{\bf Roman Cherniha,$^{a}$\footnote{\small   E-mail: r.m.cherniha@gmail.com}}
  {\bf Vasyl' Davydovych$^a$}\footnote{\small E-mail: davydovych@imath.kiev.ua}
 \\
{\it $^a$~Institute of Mathematics,  National Academy
of Sciences  of Ukraine,\\
 3, Tereshchenkivs'ka Street, Kyiv 01004, Ukraine
}\\
 \end{center}

\begin{abstract}
The well known nonlinear model for describing the solid  tumour
growth [Byrne HM., et al. Appl Math Letters 2003;16:567--74] is
under study using an approach based on Lie symmetries. It is shown
that the model
 in the two-dimensional (in space) approximation  forms  a  (1+2)-dimensional  boundary value problem,
  which admits a highly nontrivial Lie symmetry.
The special case involving the power-law nonlinearities is examined in details.
 The symmetries
derived are applied for the reduction of the nonlinear boundary
value problem  in question to  problems  of lower dimensionality.
Finally, the reduced problems with  correctly-specified coefficients
were exactly solved and  the exact solutions derived were  analysed,
in particular, some plots were build in order to understand the
time-space behaviour of these solutions and  to discuss   their
biological interpretation.

\end{abstract}

\section{\bf Introduction} \label{s1}

The Lie symmetries are widely applied to study nonlinear
differential equations (including multi-component systems of PDEs)
since 60-s of the last century, notably, for constructing their
exact solutions. There are a huge number of papers and many  books
(see, e.g.,  \cite{ovs, b-k, olv, bl-anco02, arrigo-2015,
ch-se-pl-2018})  devoted to such applications. However, one may note
that  a  small number of  them involve Lie symmetries for  solving
boundary value problems (BVPs)  based on   PDEs. To the best of our
knowledge, the first papers in this directions were published in the
beginning of 1970-s \cite{pukh-72} and \cite{bl-1974} (the extended
versions of these papers are presented in books \cite{pukh-et-al-98}
and \cite{b-k}, respectively).
 The first  book , in which an  essential role of Lie symmetries in solving
boundary value problems is discussed   and several examples are  presented, was published in 1989 \cite{rog-ames-89}.

   BVPs with moving (free) boundaries, especially those of the Stefan type,  form a special subclass  among  BVPs. They    are widely used  in mathematical  modeling  a huge number of processes, which   arise in
physics,  biology, chemistry  and  industry (see monographs
 \cite{alex93, bri-03, crank84, ready, rub71} and the papers cited therein).
 Nevertheless these processes can be very different
from formal point of view, they have  the common peculiarity,
unknown moving boundaries (free boundaries).

The classical example of    BVP with the moving boundary is the problem  modeling the ice melting.
Although  such kind of problems were  studied  earlier  by some
mathematicians (notably by   Lame and Clapeyron),\
 Jozef Stefan was the first who
 mathematically formulated, analyzed and solved  this problem. In 1889, he published four pioneering  papers
 \cite{stefan89-1,stefan89-2,stefan89-3,stefan89-4}, devoted to such kind of problems.
In order to formulate mathematically and solve analytically the ice
melting problem,   he derived  a special boundary condition
(nowadays called the Stefan conditions).  This condition reflects
the heat energy balance  at the moving boundary and
  has the form
    \[
 \quad \Gamma(t,\textbf{x})=0: \,  \lambda_{1}\frac{\partial u}{\partial \textbf{n}} = \lambda_{2}
\frac{\partial v}{\partial \textbf{n}} + L_m \textbf{U} \cdot
\textbf{n},
\] where  $u(t,\textbf{x})$ and $v(t,\textbf{x})$ are the temperatures of solid (ice)
and liquid (water) phases,  $\Gamma$ is an unknown function describing the  moving boundary, $\textbf{U}$ is  the moving boundary velocity,
   and $\textbf{n}$  is  the unit outward normal to the surface $\Gamma(t,\textbf{x})=0$ (the parameters $\lambda_{1},\ \lambda_{2}, \ L_m $ are assumed to be known positive constants with clear physical meanings, $\textbf{x}=(x_1, \dots,x_n)$. Assuming that the liquid phase temperature  $u$ is known the above condition can be rewritten in the form
    \[
 \quad \Gamma(t,\textbf{x})=0: \,   d
\frac{\partial v}{\partial \textbf{n}} = -  \textbf{U}\cdot \textbf{n} +\textbf{q}\cdot \textbf{n}, \]
which also called the Stefan condition ($\textbf{q}$ is the known function, which means the heat flux from the  solid  phase, $d= \frac{\lambda_{2}}{L_m}$).
   Because the velocity vector $\textbf{U}$ and  the normal $\textbf{n}$  can be expressed via partial derivatives of the function $\Gamma$, the condition takes the equivalent form
   (see, e.g.,
\cite{crank84}, P. 18)
    \begin{equation}\label{0-1}
   d\, \nabla v\cdot \nabla \Gamma = \Gamma_t +\textbf{q}\cdot\nabla
   \Gamma,
    \end{equation}
  where  the operator   $\nabla =
\left(\frac{\partial}{\partial x_1}, \dots,
\frac{\partial}{\partial x_n} \right)$  and the natural assumption $|\nabla \Gamma| \neq 0$
takes place.
 So, the
relevant BVP can be easily formulated by adding the standard heat
equations and the conditions on the fixed  boundary to the boundary condition (\ref{0-1}).

From the mathematical point of view,
BVPs with free  boundaries are   more complicated
objects than the standard BVPs with fixed boundaries.
In the particular case, each  BVP with
an unknown free boundary
is  nonlinear although the basic equations may be linear \cite{rub71, cr-jg59}.
 Thus, the classical methods of solving linear BVPs (the Fourier method, the method of the Laplace transformations, and so forth)  cannot  be directly applied  for solving  any BVP with free  boundaries.
However,  it can be
noted that the Lie symmetry method could be more  applicable
for solving problems with moving boundaries. In fact, the structure
of such boundaries may depend  on invariant variable(s) and this
gives a possibility to reduce the given BVP to that of lover
dimensionality. This is the reason why different  authors applied
the Lie symmetry method for solving  BVPs with free  boundaries
 \cite{ben-olv-82, bl-1974, ch-kov-09,ch-kov11a, ch-kov12, ch-da-ki-18,
pukh-06, zhou-xia-15,zhou-shi-18,tarzia1,tarzia2}. It should be
stressed that a majority of these papers are devoted to solving of
two-dimensional problems while only  a few of them are dealing with
multidimensional BVPs  \cite{ben-olv-82, ch-kov11a, pukh-06}.

In this paper,  we analyse  a  solid  tumour growth model proposed in
\cite{by-ki-2003}. The model is constructed under assumption that
the solid tumour comprises cells and water alone. The cells and
water are treated as incompressible fluids, however the tumour cells
are considered as viscous fluid while water is ideal (in-viscid)
fluid. From mathematical point of view, the model is a nonlinear BVP
with a moving  boundary. The unknown boundary  describes the tumour
growth dynamics. Because the problem is very complicated, its
one-dimensional space approximation only  was analysed   in
\cite{by-ki-2003}. Much later in \cite{ch-da-ki-18}, some exact
solutions were constructed also in the one-dimensional case and
under additional assumptions.


 Here we study this problem in the two-dimensional space approximation, i.e.
 the corresponding (1+2)-dimensional BVP, using the Lie symmetry method.
 Notably two-dimensional approximation differs essentially from one-dimensional,
  hence the results derived in our previous paper \cite{ch-da-ki-18} cannot be applied.
 In Section~\ref{s2}, Lie symmetries of the governing equations and the problem in
 question are found in two most interesting cases. In Section~\ref{s3}, a highly
 nontrivial  reduction of the given  (1+2)-dimensional BVP (with correctly-specified
 coefficients) to the problem with the  governing ODEs  is  derived. In Section~\ref{s4},
  exact solutions of the problem are constructed and analysed. In Section~\ref{s5},
  some interesting results (including exact solutions) are presented in the most general case.
 Finally, we briefly discuss the result obtained   in
the last section.

\section{\bf The model and its Lie symmetries} \label{s2}

The tumour growth model derived in \cite{by-ki-2003}  formally consists of the seven governing equations  in the
two-dimensional space approximation (see equations (3), (10)--(12) therein). However the velocity of water in the tumour can be expressed via the cell velocity using equations (11)\cite{by-ki-2003} and substituted into other equations.
Moreover the equation  for nutrients (12)\cite{by-ki-2003} may be  skipped by
treating the nutrient-rich case. As a result the system of the governing equations reduces to four PDEs.
These equations after
some simplifications (the shear viscosity coefficient
taken to be $\mu=1$ without losing generality),
 take the form
 \begin{equation}\label{2-1}
 \begin{array} {l}
  \alpha_t + \left(\alpha u^1\right)_{x}+\left(\alpha u^2\right)_{y}= S(\alpha), \medskip\\
  u^1_{x} + u^2_{y} = \nabla\cdot\left(D(\alpha)\nabla p\right),  \medskip\\
  \lf[(2+\lbd)\alpha u^1_{x} + \lbd \alpha u^2_{y}\rg]_x + \lf[\alpha u^1_{y} + \alpha u^2_{x}\rg]_y = p_x +
   (\alpha\Sigma(\alpha))_x,  \medskip\\
  \lf[\alpha u^1_{y} +  \alpha u^2_{x}\rg]_x +\lf[(2+\lbd)\alpha u^2_{y} + \lbd \alpha u^1_{x}\rg]_y = p_y + (\alpha\Sigma(\alpha))_y,
  \end{array}\end{equation}
where $\lambda>0$ is the bulk viscosity coefficient, $\alpha$ is the
tumour cell concentration, $u^1$ and $u^2$ are the cell velocity
components, $p$ is the water pressure. Here $S, \  D$  and $\Sigma$
are the known functions and they have the following meanings: $S$ is
the net cell proliferation rate, $\Sigma$ is the pressure difference
between the cell pressure and $p$,  and function $ D $ involves  the
drag coefficient $k(\alpha)$ and has the form $D=
(1-\alpha)^2/k(\alpha)$. The typical forms of the functions $k, \ S
$  and $\Sigma$  are listed in  \cite{by-ki-2003}.

Because the tumour size is changing with time, we need to supplement
the governing equations  (\ref{2-1}) by appropriate boundary
conditions. Assuming that the tumour boundary is prescribed  by a
curve $\Gamma(t,x,y)=0$, where  $\Gamma$ is an unknown function,
 the boundary conditions have the form
 \begin{equation}\label{2-2}
 \begin{array} {l}
   u^1\Gamma_{x}+ u^2\Gamma_{y}= -\Gamma_t, \quad p=0, \medskip\\
  \lf[(2+\lbd) u^1_{x} + \lbd  u^2_{y}\rg]\Gamma_x + \lf[ u^1_{y} +  u^2_{x}\rg]\Gamma_y = 0, \medskip\\
  \lf[ u^1_{y} +  u^2_{x}\rg]\Gamma_x +\lf[\lbd  u^1_{x}+(2+\lbd) u^2_{y} \rg]\Gamma_y = 0.
  \end{array}\end{equation}

  So, we have the nonlinear BVP (\ref{2-1})--(\ref{2-2}) with the  unknown moving boundary $\Gamma(t,x,y)=0$.

In order to apply Lie symmetry method for analysis of BVP
(\ref{2-1})--(\ref{2-2}), we start from description of these
symmetries of the nonlinear system  (\ref{2-1}) assuming that $D, \
S$ and   $\Sigma$ are arbitrary smooth functions.

  \begin{theo}\label{th-2-1}
  The system of nonlinear PDEs  (\ref{2-1}) with arbitrary functions $D, \ S$ and   $\Sigma$  is invariant with respect to the infinite-dimensional  Lie algebra  generated by the Lie symmetry operators
  \begin{equation}\label{2-3}
 \begin{array} {l}
   \partial_t,  \quad  F(t) \partial_p, \\
   J_f = f(t) \lf[y\partial_x -   x\partial_y +\lf(u^2+\frac{\dot f}{f}y\rg)\partial_{u^1} -
   \lf(u^1+\frac{\dot f}{f}x\rg)\partial_{u^2}\rg],\\
   G_g = g(t)\partial_x + \dot g\partial_{u^1},  \quad  G_h = h(t)\partial_y + \dot h\partial_{u^2}.
   \end{array}\end{equation}
   Here $ F, f, g, $  and $h$ are arbitrary smooth functions and the upper dot means differentiation w.r.t. time.
 \end{theo}

 \begin{remark}\label{rem-1}
Setting $ f=1$, $ g=h=1$ ,   and $g=h=t$, one easily identifies that the Galilei algebra with the basic operators
\begin{equation}\label{2-4}
 \begin{array} {l}
   \partial_t,\quad  \partial_x, \quad  \partial_y,\\
  J = y\partial_x -   x\partial_y +u^2\partial_{u^1} - u^1\partial_{u^2},\\
   G_x = t\partial_x + \partial_{u^1},  \quad  G_y = t\partial_y + \partial_{u^2}
   \end{array}\end{equation}
   is a subalgebra of (\ref{2-3}).   The Galilei algebra is the Lie   invariance algebra of many classical
   equations in  physics (see, e.g., \cite{fu-ch-89, fu-ch-95} and papers cited therein). Notably,
    representation  (\ref{2-4}) of
     this algebra coincides with that of the
      Navier--Stokes equations (in 2D space) \cite{lloyd-81}.
Thus, $J_f$ represents the generalised rotation
   symmetry and the operators $G_g$ and $G_h$ represent the generalised Galilean invariance.
 \end{remark}

Because systems (\ref{2-1}) contains three arbitrary functions, the Lie symmetry of its different
representatives depends essentially on the form of the triplet $(S, \  D,\ \Sigma)$, hence the Lie algebra of invariance can be wider than (\ref{2-3}). Thus, the problem of a complete
description of all possible Lie symmetries (the Lie symmetry  classification problem) arises. Solving this problem is a highly
 nontrivial task (see, e.g., a detailed discussion in Chapter 2 of \cite{ch-se-pl-2018}). Here we restrict ourselves to study an important special case.

It is well known that a typical time-dependence of the function
$\Gamma(t,x,y)$ is power-law. In particular, the time-dependence
$\Gamma= x- t^{1/2}$ was established by J. Stefan
 in the 1D space case for the ice melting problem mentioned above, and such
 dependence occurs in many other situations (see, e.g., the recent papers
  \cite{voller04, broa-14, ch-kov12, ch-da-ki-18,zhou-xia-15,zhou-shi-18}).
  Another typical profile is $\Gamma= x- vt$ ($v$ is an unknown velocity of
  the moving boundary), which occurs, for example, in the model describing the
  metal melting and evaporation under power energy fluxes \cite{ch93}. It means
  that  one should look for the scale invariance of the governing equations in order
  to find appropriate form of the moving boundary.

\begin{theo}\label{th-2-2}
  The system of nonlinear PDEs  (\ref{2-1}) with the  functions
   \begin{equation}\label{2-5}
  D=d_0\al^m, \ S =s_0\al^n, \ \Sigma=\sigma_0\al^{n-1}
  \end{equation}
  is invariant  under the scale transformation generated by the
   Lie symmetry operator
   \begin{equation}\label{2-6}
   D_{mn}= 2(1-n)t\p_t +(1+m)(x\partial_x +y\partial_y)+2\al\p_{\al} + (m+2n-1)(u^1\partial_{u^1} +
   u^2\partial_{u^2}) + 2np\p_p\,,
   \end{equation}
   where $d_0, \ s_0, \ \sigma_0,\ m$  and $n$ are arbitrary constants.
   \end{theo}

 \begin{remark}\label{rem-1*}
 The power-law profile of the function $D=(1-\alpha)^2/k(\alpha)$  is natural because one corresponds to
 the drag coefficient   $k(\alpha)=d_0^{-1}\al^{-m}(1-\alpha)^2$.  Obviously, this function   with $m<0$ is a
  natural generalisation of the logistic profile $\al(1-\alpha)$ proposed
  in  \cite{by-ki-2003} and  the case $m=-1$ is the most interesting. Moreover,
   this exponent   naturally  stands out from others in Section~\ref{s4}.
  \end{remark}

Let us examine the Lie symmetry of the nonlinear BVP
(\ref{2-1})--(\ref{2-2}) using the definition proposed in
\cite{ch-kov11a} and assuming $\Gamma$ to be a closed curve for any
$t\geq 0$. If the boundary $\Gamma(t,x, y)=0$ contains points at
infinity (e.g., it is  a strip moving in time) then a generalisation
of the definition proposed in \cite{ch-king-15} is needed.

\begin{theo}\label{th-2-3}
{ \it (i)} The  nonlinear BVP (\ref{2-1})--(\ref{2-2}) admits the
Lie symmetry operator $J_f$ from (\ref{2-3}),
  provided the moving boundary has the circular form  $\Gamma(t,x^2+y^2)=0$.\\
  { \it (ii)} The  nonlinear BVP (\ref{2-1})--(\ref{2-2}) admits the Lie symmetry operators $J_f$ from (\ref{2-3})
 and $ D_{mn}$ (\ref{2-6}) with $n\neq1$
  provided the functions $D, \ S $  and $\Sigma$ have the forms (\ref{2-5})
  and  the (circular) moving boundary is specified as follows
   \begin{equation}\label{2-7*} \Gamma\Big(\frac{x^2+y^2}{t^\kappa}\Big)=0,  \quad \kappa=\frac{1+m}{1-n}.  \end{equation}
     \end{theo}

\textbf{The sketch of the proof.} Here we use  Definition~2
\cite{ch-kov11a} and  consider BVP in question as a system of
manifolds in prolonged spaces. We need to show that each manifold is
invariant w.r.t. the  Lie group generated by the operators $J_f$ and
$ D_{mn}$. Obviously that there is no need to examine manifolds
corresponding to the governing equations because of
Theorems~\ref{th-2-1} and \ref{th-2-2}. So, we need to show only
invariance of the manifolds corresponding to the boundary
conditions~(\ref{2-2}).

 Let us consider case {\it (i)} in detail.
The Lie group generated by the operator $J_f$ has the
form:
\begin{equation}\label{2-7}
 \begin{array} {l}
 \medskip
t^{\ast}=t, \ x^{\ast} = x\cos \left(f\varepsilon\right) + y \sin
\left(f\varepsilon\right), \
y^{\ast} = -x\sin\left(f\varepsilon\right) + y \cos\left(f\varepsilon\right),\\
u^{1^{\ast}}= \left(u^1+\dot{f}\varepsilon
y\right)\cos\left(f\varepsilon\right) + \left(u^2-\dot{f}\varepsilon
x\right) \sin\left(f\varepsilon\right), \\
 u^{2^{\ast}} =
-\left(u^1+\dot{f}\varepsilon y\right)\sin\left(f\varepsilon\right)
+
\left(u^2-\dot{f}\varepsilon x\right) \cos\left(f\varepsilon\right), \\
\alpha^{\ast}=\alpha, \ p^{\ast}=p,
 \end{array}\end{equation}
where $\varepsilon$ is the group parameter. Formally speaking it is a infinite-dimensional group but one may consider the function $f(t)$ as a  parameter because the time $t$ is unchanged under action of  (\ref{2-7}). Hence  the Lie group (\ref{2-7}) acts like the standard rotation group corresponding to the operator $J$ from (\ref{2-4}), excepting the manifolds involving time-derivatives (see the first equation  in (\ref{2-2})).

Since the BVP (\ref{2-1})--(\ref{2-2})  has free boundary
$\Gamma(t,x,y)=0$, we need also to extend the group (\ref{2-7})  by
adding transformation for the new variable $\Gamma^{\ast}=\Gamma$ (according to Definition~2 \cite{ch-kov11a}).

To prove  the invariance of the boundary condition (\ref{2-2}) with
respect to the Lie group (\ref{2-7}), we need to find
transformations  for  the variables $u^1,\ u^2, \ \Gamma$ and their
first-order  derivatives. As a result, the following  formulae  were
derived:
\begin{equation}\label{2-8}
 \begin{array} {l} \medskip
 u^1=u^{1^{\ast}}\cos\left(f\varepsilon\right)-u^{2^{\ast}}\sin\left(f\varepsilon\right)-\dot{f}\varepsilon\, y,\\
 \medskip
 u^2=u^{1^{\ast}}\sin\left(f\varepsilon\right)+u^{2^{\ast}}\cos\left(f\varepsilon\right)+\dot{f}\varepsilon\, x,\\
 \medskip
u^1_x=\cos^2\left(f\varepsilon\right)u^{1^{\ast}}_{\ x^{\ast}}+
\sin^2\left(f\varepsilon\right)u^{2^{\ast}}_{\ y^{\ast}}-\sin
\left(f\varepsilon\right) \cos
\left(f\varepsilon\right)\left(u^{1^{\ast}}_{\ y^{\ast}}+u^{2^{\ast}}_{\ x^{\ast}}\right),\\
\medskip u^1_y=\cos^2\left(f\varepsilon\right)u^{1^{\ast}}_{\ y^{\ast}}-
\sin^2\left(f\varepsilon\right)u^{2^{\ast}}_{\ x^{\ast}}+\sin
\left(f\varepsilon\right) \cos
\left(f\varepsilon\right)\left(u^{1^{\ast}}_{\ x^{\ast}}-u^{2^{\ast}}_{\ y^{\ast}}\right)-\dot{f}\varepsilon,\\
\medskip u^2_x=-\sin^2\left(f\varepsilon\right)u^{1^{\ast}}_{\ y^{\ast}}+
\cos^2\left(f\varepsilon\right)u^{2^{\ast}}_{\ x^{\ast}}+\sin
\left(f\varepsilon\right) \cos
\left(f\varepsilon\right)\left(u^{1^{\ast}}_{\ x^{\ast}}-u^{2^{\ast}}_{\ y^{\ast}}\right)+\dot{f}\varepsilon,\\
\medskip
 u^2_y=\sin^2\left(f\varepsilon\right)u^{1^{\ast}}_{\ x^{\ast}}+
\cos^2\left(f\varepsilon\right)u^{2^{\ast}}_{\ y^{\ast}}+\sin
\left(f\varepsilon\right) \cos
\left(f\varepsilon\right)\left(u^{1^{\ast}}_{\ y^{\ast}}+u^{2^{\ast}}_{\ x^{\ast}}\right),\\
\medskip \Gamma_x=\cos \left(f\varepsilon\right)\Gamma^{\ast}_{x^{\ast}}-\sin
\left(f\varepsilon\right)\Gamma^{\ast}_{y^{\ast}}, \ \Gamma_y=\sin
\left(f\varepsilon\right)\Gamma^{\ast}_{x^{\ast}}+\cos
\left(f\varepsilon\right)\Gamma^{\ast}_{y^{\ast}},
\\ \medskip
\Gamma_t=\Gamma^{\ast}_{t^{\ast}}+\dot{f}\varepsilon\left(-x\sin\left(f\varepsilon\right)
+ y \cos\left(f\varepsilon\right)\right)\Gamma^{\ast}_{
x^{\ast}}-\dot{f}\varepsilon\left(x\cos \left(f\varepsilon\right) +
y \sin \left(f\varepsilon\right)\right)\Gamma^{\ast}_{ y^{\ast}}.
\end{array}\end{equation}

Substituting (\ref{2-8}) into (\ref{2-2}) and making straightforward
calculations, we arrive at the equations:
 \begin{equation}\label{2-9}
 \begin{array} {l}
   u^{1^{\ast}}\Gamma^{\ast}_{ x^{\ast}}+ u^{2^{\ast}}\Gamma^{\ast}_{ y^{\ast}}= -\Gamma^{\ast}_{t^{\ast}}, \quad p^{\ast}=0, \medskip\\
  \lf[-\sin\left(f\varepsilon\right)\left(u^{1^{\ast}}_{\ y^{\ast}} +
  u^{2^{\ast}}_{\ x^{\ast}}\right)+\cos\left(f\varepsilon\right)\left((2+\lambda)u^{1^{\ast}}_{\ x^{\ast}} +
  \lambda u^{2^{\ast}}_{\ y^{\ast}}\right)\rg]\Gamma^{\ast}_{ x^{\ast}}
  +\medskip\\ \hskip1cm
  \lf[\cos\left(f\varepsilon\right)\left(u^{1^{\ast}}_{\ y^{\ast}} +
  u^{2^{\ast}}_{\ x^{\ast}}\right)-\sin\left(f\varepsilon\right)\left( \lambda u^{1^{\ast}}_{\ x^{\ast}} +
  (2+\lambda)u^{2^{\ast}}_{\ y^{\ast}}\right)\rg]\Gamma^{\ast}_{ y^{\ast}} = 0, \medskip\\
   \lf[\cos\left(f\varepsilon\right)\left(u^{1^{\ast}}_{\ y^{\ast}} +
  u^{2^{\ast}}_{\ x^{\ast}}\right)+\sin\left(f\varepsilon\right)\left((2+\lambda)u^{1^{\ast}}_{\ x^{\ast}} +
  \lambda u^{2^{\ast}}_{\ y^{\ast}}\right)\rg]\Gamma^{\ast}_{ x^{\ast}}
  +\medskip\\ \hskip1cm
  \lf[\sin\left(f\varepsilon\right)\left(u^{1^{\ast}}_{\ y^{\ast}} +
  u^{2^{\ast}}_{\ x^{\ast}}\right)+\cos\left(f\varepsilon\right)\left( \lambda u^{1^{\ast}}_{\ x^{\ast}} +
  (2+\lambda)u^{2^{\ast}}_{\ y^{\ast}}\right)\rg]\Gamma^{\ast}_{ y^{\ast}} =
  0.
  \end{array}\end{equation}
Using the linear combinations of the last two equations, system
(\ref{2-9}) can be rewritten as
  \begin{equation}\label{2-9*}
 \begin{array} {l}
   u^{1^{\ast}}\Gamma^{\ast}_{ x^{\ast}}+ u^{2^{\ast}}\Gamma^{\ast}_{ y^{\ast}}= -\Gamma^{\ast}_{t^{\ast}}, \quad p^{\ast}=0, \medskip\\
 \lf[(2+\lambda)u^{1^{\ast}}_{\ x^{\ast}} +
  \lambda u^{2^{\ast}}_{\ y^{\ast}}\rg]\Gamma^{\ast}_{ x^{\ast}}
  +  \lf[u^{1^{\ast}}_{\ y^{\ast}} +
  u^{2^{\ast}}_{\ x^{\ast}}\rg]\Gamma^{\ast}_{ y^{\ast}} = 0, \medskip\\
   \lf[u^{1^{\ast}}_{\ y^{\ast}} +
  u^{2^{\ast}}_{\ x^{\ast}}\rg]\Gamma^{\ast}_{x^{\ast}}
  +
  \lf[ \lambda u^{1^{\ast}}_{\ x^{\ast}} +
  (2+\lambda)u^{2^{\ast}}_{\ y^{\ast}}\rg]\Gamma^{\ast}_{ y^{\ast}} =
  0.
  \end{array}\end{equation}
Thus,  the  Lie group (\ref{2-7}) transforms
boundary conditions (\ref{2-2}) to the same form (\ref{2-9*}). This
means that conditions (\ref{2-2})
 are invariant w.r.t. the operator $J_f$.

 To complete the proof of case {\it (i)}, we need to show invariance of  the moving boundary $\Gamma(t,x,y)=0$. Because the Lie group (\ref{2-7}) acts like the  rotation group on $x$ and $y$, we immediately conclude that the moving boundary has the form
    \be\label{2-10}\Gamma(t,x^2+y^2)=0.\ee

Case {\it (ii)} of the theorem can be proved in  a quite similar way, i.e.   it  can be  shown that the boundary
conditions (\ref{2-2}) are invariant w.r.t. the scale transformations generated
 by the operator $D_{mn}$.
 Notably, the form of the function $\Gamma$ can be established also by using the standard Lie invariance criteria:
\be\label{2-11} D_{mn} (\Gamma)\Big\vert_{\Gamma=0}=0.\ee Taking
into account  (\ref{2-10}), equation (\ref{2-11}) leads to
\be\nonumber
(1-n)t\,\Gamma_t+(1+m)\omega\,\Gamma_\omega\Big\vert_{\Gamma(t,\omega)=0}=0\ee
(here $\omega=x^2+y^2$) that immediately gives (\ref{2-7*}).

Thus, the sketch of the proof is now
 complete.

\section{\bf Reduction of the boundary-value problem
(\ref{2-1})--(\ref{2-2})} \label{s3}

A typical time-dependence of unknown boundary of BVP with moving
boundaries is power-law. So, if  the nonlinear BVP in question
is invariant under scale transformations then this guarantees the needed form of the moving boundary.
 As it follows from
Theorem~\ref{th-2-3}, the governing equations (\ref{2-1}) should be
\begin{equation}\label{2-1*}
 \begin{array} {l}
  \alpha_t + \left(\alpha u^1\right)_{x}+\left(\alpha u^2\right)_{y}= s_0\al^n,\\
  u^1_{x} + u^2_{y} = d_0\,\nabla\cdot\left(\al^m\nabla p\right), \\
  \lf[(2+\lbd)\alpha u^1_{x} + \lbd \alpha u^2_{y}\rg]_x + \lf[\alpha u^1_{y} + \alpha u^2_{x}\rg]_y = p_x +
   \sigma_0\lf(\al^{n}\rg)_x, \\
  \lf[\alpha u^1_{y} +  \alpha u^2_{x}\rg]_x +\lf[(2+\lbd)\alpha u^2_{y} + \lbd \alpha u^1_{x}\rg]_y = p_y +
     \sigma_0\lf(\al^{n}\rg)_y,
  \end{array}\end{equation}
  and the function $\Gamma$ must have  the form (\ref{2-7*}).

The ansatz corresponding to the operator of scale transformations
(\ref{2-6}) can be easily derived, namely:
 \be\label{3-1}\ba
u^1=t^{-\gamma-1}U^1(\omega_1,\omega_2), \
u^2=t^{-\gamma-1}U^2(\omega_1,\omega_2), \\
\alpha=t^{\frac{1}{1-n}}\Lambda(\omega_1,\omega_2), \
p=t^{\frac{n}{1-n}}P(\omega_1,\omega_2), \ea\ee where
$\omega_1=xt^{\gamma}$ and $\omega_2=yt^{\gamma}$
 are new invariant
variables ($\gamma=\frac{m+1}{2(n-1)}$  and $n\not=1$),  while the
capital letters in RHS denote new unknown functions.

Substituting the ansatz (\ref{3-1}) into (\ref{2-1*}), one obtains
the reduced system of PDEs \be\label{3-4}\ba
\gamma\lf(\omega_1\Lambda_{\omega_1}+\omega_2\Lambda_{\omega_2}\rg)+\lf(\Lambda
U^1\rg)_{\omega_1}+
\lf(\Lambda U^2\rg)_{\omega_2}=s_0\Lambda^n+\frac{\Lambda}{n-1},\medskip\\
U^1_{\omega_1}+U^2_{\omega_2}=d_0\Lambda^m\lf(P_{\omega_1\omega_1}+P_{\omega_2\omega_2}\rg)+d_0m\Lambda^{m-1}
\lf(\Lambda_{\omega_1}P_{\omega_1}+\Lambda_{\omega_2}P_{\omega_2}\rg),\medskip\\
(2+\lambda)\lf(\Lambda
U^1_{\omega_1}\rg)_{\omega_1}+\lambda\lf(\Lambda
U^2_{\omega_2}\rg)_{\omega_1}+
\lf(\Lambda\lf(U^1_{\omega_2}+U^2_{\omega_1}\rg)\rg)_{\omega_2}=P_{\omega_1}+\sigma_0\lf(\Lambda^n\rg)_{\omega_1},\medskip\\
(2+\lambda)\lf(\Lambda
U^2_{\omega_2}\rg)_{\omega_2}+\lambda\lf(\Lambda
U^1_{\omega_1}\rg)_{\omega_2}+
\lf(\Lambda\lf(U^1_{\omega_2}+U^2_{\omega_1}\rg)\rg)_{\omega_1}=P_{\omega_2}+\sigma_0\lf(\Lambda^n\rg)_{\omega_2}.\ea\ee

The moving boundary $\Gamma(t,x,y)=0$ takes the form
\be\label{3-2}\Gamma\equiv\omega_1g\lf(\frac{\omega_2}{\omega_1}\rg)-1=0.\ee

Substituting the ansatz (\ref{3-1}) into (\ref{2-2}) and taking into
account (\ref{3-2}), we obtain the reduced boundary conditions at
$\Gamma =0$\,:
\be\label{3-3}\ba \frac{dg}{d\omega}\lf(U^2-\frac{\omega_2}{\omega_1}\,U^1\rg)+\frac{1}{\omega_1}\,U^1+\gamma=0, \ P=0,\medskip\\
\lf(1-\omega_2\,\frac{dg}{d\omega}\rg)\lf((2+\lambda)U^1_{\omega_1}+\lambda\,
U^2_{\omega_2}\rg)+\omega_1\frac{dg}{d\omega}\lf(U^1_{\omega_2}+U^2_{\omega_1}\rg)=0,\medskip\\
\lf(1-\omega_2\,\frac{dg}{d\omega}\rg)\lf(U^1_{\omega_2}+U^2_{\omega_1}\rg)
+\omega_1\frac{dg}{d\omega}\lf((2+\lambda)U^2_{\omega_2}+\lambda\,
U^1_{\omega_1}\rg)=0, \ea\ee where
$\omega\equiv\frac{\omega_2}{\omega_1}.$

It can be shown using the definition from \cite{ch-kov11a} and using
the algorithm presented above for the proof of Theorem~\ref{th-2-3}
that the BVP (\ref{3-4}) and (\ref{3-3}) also possesses nontrivial
Lie symmetries.

\begin{theo}\label{th-2-4} The nonlinear BVP
(\ref{3-4}), (\ref{3-3}) is invariant w.r.t. the three-dimensional
Lie algebra with the basic operators  \be\label{3-5}
J=\omega_2\p_{\omega_1}-\omega_1\p_{\omega_2}+U^2\p_{U^1}-U^1\p_{U^2},\ee
\be\nonumber \p_{\omega_1}-\frac{m+1}{2(n-1)}\p_{U^1}, \quad
\p_{\omega_2}-\frac{m+1}{2(n-1)}\p_{U^2}. \ee

\end{theo}


It should be stressed that it is rather unusual that the reduced BVP
possesses a nontrivial symmetry. For example the reduced problem
derived in \cite{ch-kov11a} (see formulae (56)--(60)) for a
multidimensional BVP describing the metal melting and evaporation  does not allow any
nontrivial Lie symmetry, therefore a non-Lie ansatz was applied for
the further reduction. Here we can make the reduction of
 the two-dimensional  BVP (\ref{3-4}), (\ref{3-3}) using the
Lie symmetry operator (\ref{3-5}).

In order to simplify calculations, we rewrite the nonlinear BVP
(\ref{3-4}) and (\ref{3-3}) in the polar coordinates applying the
formulae
 \be\label{3-6}\begin{array}{l} \omega_1=r\cos\phi, \  \omega_2=r\sin\phi,\\
 U^1=R(r,\phi)\cos\Phi(r,\phi),\ U^2=R(r,\phi)\sin\Phi(r,\phi),\\
 \Lambda=\Lambda(r,\phi), \ P=P(r,\phi).
 \end{array}\ee
Obviously, formulae (\ref{3-6}) transforms operator (\ref{3-5}) to
the form \be\label{3-7}J=-\p_{\phi}-\p_{\Phi}.\ee The ansatz
corresponding to operator (\ref{3-7}) can be easily derived, namely:
\be\label{3-8}\ba R=R_*(r), \ \Phi=\Phi_*(r)+\phi, \
\Lambda=\Lambda_*(r), \ P=P_*(r), \ea\ee where the letters with
lower stars denote   new unknown functions.

Thus, substituting (\ref{3-6}) and (\ref{3-8}) into BVP
(\ref{3-4})--(\ref{3-3}) and omitting the relevant calculations, we obtain a BVP with the governing
equations  \be\label{3-9}\ba \frac{m+1}{2(n-1)}\,
r^2\Lambda_*'+\lf(r\Lambda_*R_*\cos\Phi_*\rg)'=s_0r\Lambda_*^n+\frac{r}{n-1}\,\Lambda_*,\medskip\\
\lf(rR_*\cos\Phi_*\rg)'=d_0\lf(r\Lambda_*^{m}P_*'\rg)',\medskip\\
(1+\lambda)R_* \Lambda_*'\sin2\Phi_*-(2+\lambda)\lf(rR_*
\Lambda_*\Phi_*'\rg)'-(2+\lambda)r\Lambda_*R_*'\Phi_*'
=\\ r\lf(\sigma_0\lf(\Lambda_*^n\rg)'+
P_*'\rg)\sin\Phi_*,\medskip\\
(1+\lambda)rR_*
\Lambda_*'\cos2\Phi_*+(2+\lambda)r\lf(r\Lambda_*R_*'\rg)'-(2+\lambda)\Lambda_*R_*\lf(1+r^2\Phi_*'^2\rg)-rR_*
\Lambda_*'
=\\r^2\lf(\sigma_0\lf(\Lambda_*^n\rg)'+P_*'\rg)\cos\Phi_*,\ea\ee
where the upper prime means differentiation w.r.t. the variable $r$.

In order to reduce  the boundary conditions (\ref{3-3}), one need to
specify the function $\Gamma$ in (\ref{3-2}). Rewriting  $\Gamma$ in
polar coordinates, we immediately obtain  that (\ref{3-2}) is
invariant under the operator (\ref{3-7}) provided
\be\nonumber\Gamma\equiv \frac{r}{\delta}-1=0,\ee where $\delta>0$
is an arbitrary constant at the moment.  So, the boundary conditions
(\ref{3-3}) are reduced to
  \be\label{3-10}\ba r=\delta: \quad \frac{m+1}{2(n-1)}\,r+R_*\cos\Phi_*=0, \ P_*=0,\medskip\\
r=\delta: \quad (2+\lambda)rR_*'+R_*\lf((1+\lambda)\cos2\Phi_*-1\rg)=0,\medskip\\
r=\delta: \quad
R_*\lf((2+\lambda)r\Phi_*'-(1+\lambda)\sin2\Phi_*\rg)=0.\ea\ee
As a result,  BVP (\ref{3-9})--(\ref{3-10}) is derived.
Notably, the governing equations of this BVP are ODEs (not PDEs).

\section{\bf Exact solutions of the  boundary-value problem
(\ref{2-1})--(\ref{2-2})} \label{s4}

The nonlinear BVP (\ref{3-9})--(\ref{3-10}) is still  a complicated
problem and we were unable to solve it  in the general  case.
Happily, we were able to derive  exact solutions under additional
correctly-specified restrictions.

In fact, the system (\ref{3-9}) contains both power-law
nonlinearities and trigonometric functions.  In order to have only
 power-law  nonlinearities,  we put
\be\label{4-0}\sin\Phi_*=0 \ee
(notably the assumption $\cos\Phi_*=0$  does not lead to any interesting results).  In this case,  one immediately
obtains \be\label{4-1} R_*=\pm
\lf(\frac{\beta}{r}+d_0\Lambda_*^mP_*'\rg)\ee (hereafter  $\beta$ is an
arbitrary constant) from the second equation of (\ref{3-9}), while
the third equation of (\ref{3-9}) is satisfied identically. In
(\ref{4-1}), the sign `$+$' corresponds to the value $\Phi_*=2k\pi$,
while the sign `$-$' corresponds to the value $\Phi_*=(2k+1)\pi,$
$k\in\mathbf{Z}$.
It can be shown that one may set the sign `$+$' without losing the
generality, i.e. \be\label{4-2}
R_*=\frac{\beta}{r}+d_0\Lambda_*^mP_*'.\ee

\begin{remark} \label{rem-3}
From the physical point of view, formulae (\ref{4-0}) and
(\ref{4-1}) mean that a generalization of the  classical radially
symmetric flow takes place. In the case of a constant pressure, we
obtain the radially symmetric flow $R_*=\frac{\beta}{r}$, where
$\beta$  is proportional of the constant rate of fluid, which is
supplied from a source (or to a sink) in point $(0;0)$.
\end{remark}

Substituting (\ref{4-2}) into the first and fourth equations of
(\ref{3-9}), we obtain the nonlinear ODE system with respect to the
functions $\Lambda_*$ and $P_*$ of the form \be\label{4-3}\ba
\lf(r\Lambda_*^{1+m}P_*'\rg)'+
\frac{(1+m)r^2+2(n-1)\beta}{2d_0(n-1)}\,\Lambda_*'-\frac{s_0}{d_0}\,r\Lambda_*^{n}-\frac{1}{d_0(n-1)}\,r\Lambda_*=0,\medskip\\
\Lambda_*^{1+m}P_*'''+\Lambda_*^{m}\lf(\frac{\Lambda_*}{r}+(1+2m)\Lambda_*'\rg)P_*''+
m\Lambda_*^{m}P_*'\Lambda_*''+m^2\Lambda_*^{m-1}P_*'{\Lambda_*'}^2+\medskip\\
\lf(\frac{\lambda}{(2+\lambda)r}+\frac{m}{r}\rg)\Lambda_*^{m}P_*'\Lambda_*'-\lf(\frac{1}{d_0(2+\lambda)}+
\frac{\Lambda_*^{1+m}}{r^2}\rg)P_*'
-\lf(\frac{2\beta}{d_0(2+\lambda)r^2}+\frac{n\sigma_0\Lambda_*^{n-1}}{d_0(2+\lambda)}\rg)\Lambda_*'
=0.\ea\ee

Using the first equation of (\ref{4-3}) and its differential
consequences with respect to $r$, one can find the expressions for
 $P_*''$ and $P_*'''$. Substituting
the expressions obtained into the second equation of (\ref{4-3}), we
arrive at the nonlinear ODE  \be\label{4-4}\ba
2(n-1)d_0\lf(\Lambda_*^{m}\Lambda_*''-\Lambda_*^{m-1}{\Lambda_*'}^2-
\frac{\lambda}{(2+\lambda)r}\,\Lambda_*^{m}\Lambda_*'+\frac{1}{d_0(2+\lambda)}\rg)P_*'+\\
\lf((1+m)r+\frac{2(n-1)\beta}{r}\rg)\lf(\Lambda_*''-\frac{{\Lambda_*'}^2}{\Lambda_*}\rg)+
2(n-1)\lf(\frac{n\sigma_0}{2+\lambda}-(n-1)s_0\rg)\Lambda_*^{n-1}\Lambda_*'+\\
\lf(1+m-\frac{2(n-1)\beta\lambda}{(2+\lambda)r^2}\rg)\Lambda_*'=0.\ea\ee

Thus, to find the functions $P_*$ and $\Lambda_*$  we must to solve
the nonlinear system consisting of the first equation in (\ref{4-3})
and equation (\ref{4-4}). The construction of
   the general solution of this system is a difficult task.
Hence we look for a particular solution assuming that equation
(\ref{4-4}) is satisfied identically for each  function $P_*$. Thus,
the overdetermined system
 \be\label{4-5}\ba
\Lambda_*^{m}\Lambda_*''-\Lambda_*^{m-1}{\Lambda_*'}^2-
\frac{\lambda}{(2+\lambda)r}\,\Lambda_*^{m}\Lambda_*'+\frac{1}{d_0(2+\lambda)}=0,\\
\lf((1+m)r+\frac{2(n-1)\beta}{r}\rg)\lf(\Lambda_*''-\frac{{\Lambda_*'}^2}{\Lambda_*}\rg)+
2(n-1)\lf(\frac{n\sigma_0}{2+\lambda}-(n-1)s_0\rg)\Lambda_*^{n-1}\Lambda_*'+\\
\lf(1+m-\frac{2(n-1)\beta\lambda}{(2+\lambda)r^2}\rg)\Lambda_*'=0\ea\ee
needs to be solved.

A linear combination of equations (\ref{4-5}) leads to the nonlinear
equation \be\label{4-6}\ba
\Big((1+m)(1+\lambda)\Lambda_*^{m}+(n-1)\big(n\sigma_0-(n-1)(2+\lambda)s_0\big)\Lambda_*^{m+n-1}\Big)\Lambda_*'=
\frac{1+m}{2d_0}\,r+\frac{(n-1)\beta}{d_0r}\,,\ea\ee which can be
easily
 integrated. Using the general solution of (\ref{4-6}), it is
  possible to reduce the first equation in (\ref{4-5}) to an algebraic equation for the function $\Lambda_*$.
  As a result, exactly two  forms of $\Lambda_*$ were derived, namely
\be\label{4-7}\Lambda_*=c_1\exp\lf(c_2r^{\frac{2+2\lambda}{2+\lambda}}-\frac{r^2}{4d_0}\rg),\ee
if $m=-1, \ \beta=0, \ s_0=\frac{n\sigma_0}{(n-1)(2+\lambda)}\,,$ \\
and
\be\label{4-8}\Lambda_*=c_1r^{\frac{2}{1+m}},\ee if
 $m\neq-1, \ \beta=0, \ s_0=\frac{n\sigma_0}{(n-1)(2+\lambda)}, \
d_0=\frac{1+m}{4(1+\lambda)c_1^{1+m}}$  (hereafter $c_i\ (i=1,2,
\dots)$ are arbitrary constants).

Let us consider the function  $\Lambda_*$ from (\ref{4-7}) in detail
for exact  solving   the nonlinear   BVP (\ref{3-9}) and
(\ref{3-10}). Substituting  $\Lambda_*$ into the first equation in
(\ref{4-3}), we derive the linear ODE \be\nonumber\lf(rP_*'\rg)'
=\frac{n\sigma_0c_1^n}{d_0(n-1)(2+\lambda)}\,r\exp\lf(c_2n\,r^{\frac{2+2\lambda}{2+\lambda}}-\frac{nr^2}{4d_0}\rg)
+\frac{c_1}{d_0(n-1)}\,r\exp\lf(c_2r^{\frac{2+2\lambda}{2+\lambda}}-\frac{r^2}{4d_0}\rg),\ee
hence  its general solution is
 \be\label{4-9}\ba P_*=c_4+c_3\ln r+\frac{n\sigma_0c_1^n}{d_0(n-1)(2+
 \lambda)}\,\int \frac{\int r\exp\lf(c_2n\,r^{\frac{2+2\lambda}{2+\lambda}}-\frac{nr^2}{4d_0}\rg)
 dr}{r}\,dr+\medskip\\ \hskip3cm \frac{c_1}{d_0(n-1)}\int\frac{\int \,r\exp\lf(c_2r^{\frac{2+
 2\lambda}{2+\lambda}}-\frac{r^2}{4d_0}\rg) dr}{r}\,dr.\ea\ee
In the case $c_2=0$, the solution (\ref{4-9}) can be essentially
simplified to  \be\label{4-10}\ba P_*=c_4+c_3\ln
r+\frac{2\sigma_0c_1^n}{(n-1)(2+
 \lambda)}\,\int^\delta_r \frac{\exp\lf(-\frac{nz^2}{4d_0}\rg)}{z}\,dz+
 \frac{2c_1}{n-1}\int^\delta_r\frac{\exp\lf(-\frac{z^2}{4d_0}\rg)}{z}\,dz,\ea\ee
where $r<\delta$ and the constant $\delta$ should be specified using
the boundary conditions (\ref{3-10}).

Thus, using formulae (\ref{4-0}), (\ref{4-2}), (\ref{4-7}) and
(\ref{4-10}), we obtain the exact solution \be\label{4-11}\ba
\Phi_*=2k\pi, \ \Lambda_*=c_1\exp\lf(-\frac{r^2}{4d_0}\rg),\\
P_*=c_4+c_3\ln r+\frac{2\sigma_0c_1^n}{(n-1)(2+
 \lambda)}\,\int^\delta_r \frac{\exp\lf(-\frac{nz^2}{4d_0}\rg)}{z}\,dz+
 \frac{2c_1}{n-1}\int^\delta_r\frac{\exp\lf(-\frac{z^2}{4d_0}\rg)}{z}\,dz,\\
 R_*=\frac{c_3d_0}{c_1r}\,\exp\lf(\frac{r^2}{4d_0}\rg)-
\frac{2\sigma_0d_0c_1^{n-1}}{(n-1)(2+
 \lambda)r}\,\exp\lf(\frac{(1-n)r^2}{4d_0}\rg)-\frac{2d_0}{(n-1)r}\ea\ee
 of the ODE system (\ref{3-9}) with $m=-1$ and $s_0=\frac{n\sigma_0}{(n-1)(2+\lambda)}.$

Substituting result obtained  into (\ref{3-8}), (\ref{3-6}) and
(\ref{3-1}), we  find the exact solution \be\label{4-13}\ba
u^1=\frac{d_0x}{t(x^2+y^2)}\lf[\frac{c_3}{c_1}\,\exp\lf(\frac{x^2+y^2}{4d_0}\rg)-
\frac{2\sigma_0c_1^{n-1}}{(n-1)(2+
 \lambda)}\,\exp\lf(\frac{(1-n)(x^2+y^2)}{4d_0}\rg)-\frac{2}{(n-1)}\rg], \medskip
\\ u^2=\frac{d_0y}{t(x^2+y^2)}\lf[\frac{c_3}{c_1}\,\exp\lf(\frac{x^2+y^2}{4d_0}\rg)-
\frac{2\sigma_0c_1^{n-1}}{(n-1)(2+
 \lambda)}\,\exp\lf(\frac{(1-n)(x^2+y^2)}{4d_0}\rg)-\frac{2}{(n-1)}\rg], \medskip
\\
p=t^\frac{n}{1-n}\lf[\frac{2\sigma_0c_1^n}{(n-1)(2+
 \lambda)}\,\mbox{\raisebox{-3.2ex}{$\stackrel{\displaystyle\int^\delta}{\scriptstyle \sqrt{x^2+y^2}\hskip0.2cm}$}}
 \frac{\exp\lf(-\frac{nz^2}{4d_0}\rg)}{z}\,dz+ \rg.\\
\hskip1cm + \lf.
\frac{2c_1}{n-1}\mbox{\raisebox{-3.2ex}{$\stackrel{\displaystyle\int^\delta}{\scriptstyle
\sqrt{x^2+
y^2}\hskip0.2cm}$}}\frac{\exp\lf(-\frac{z^2}{4d_0}\rg)}{z}\,dz+c_4+\frac{c_3}{2}\ln (x^2+y^2)\rg],\\
\alpha=c_1t^\frac{1}{1-n}\exp\lf(-\frac{x^2+y^2}{4d_0}\rg) \\
 \ea\ee of the nonlinear system (\ref{2-1*}) with $m=-1$ and
 $s_0=\frac{n\sigma_0}{(n-1)(2+\lambda)}.$

\begin{figure}
\centering
\includegraphics[width=7 cm]{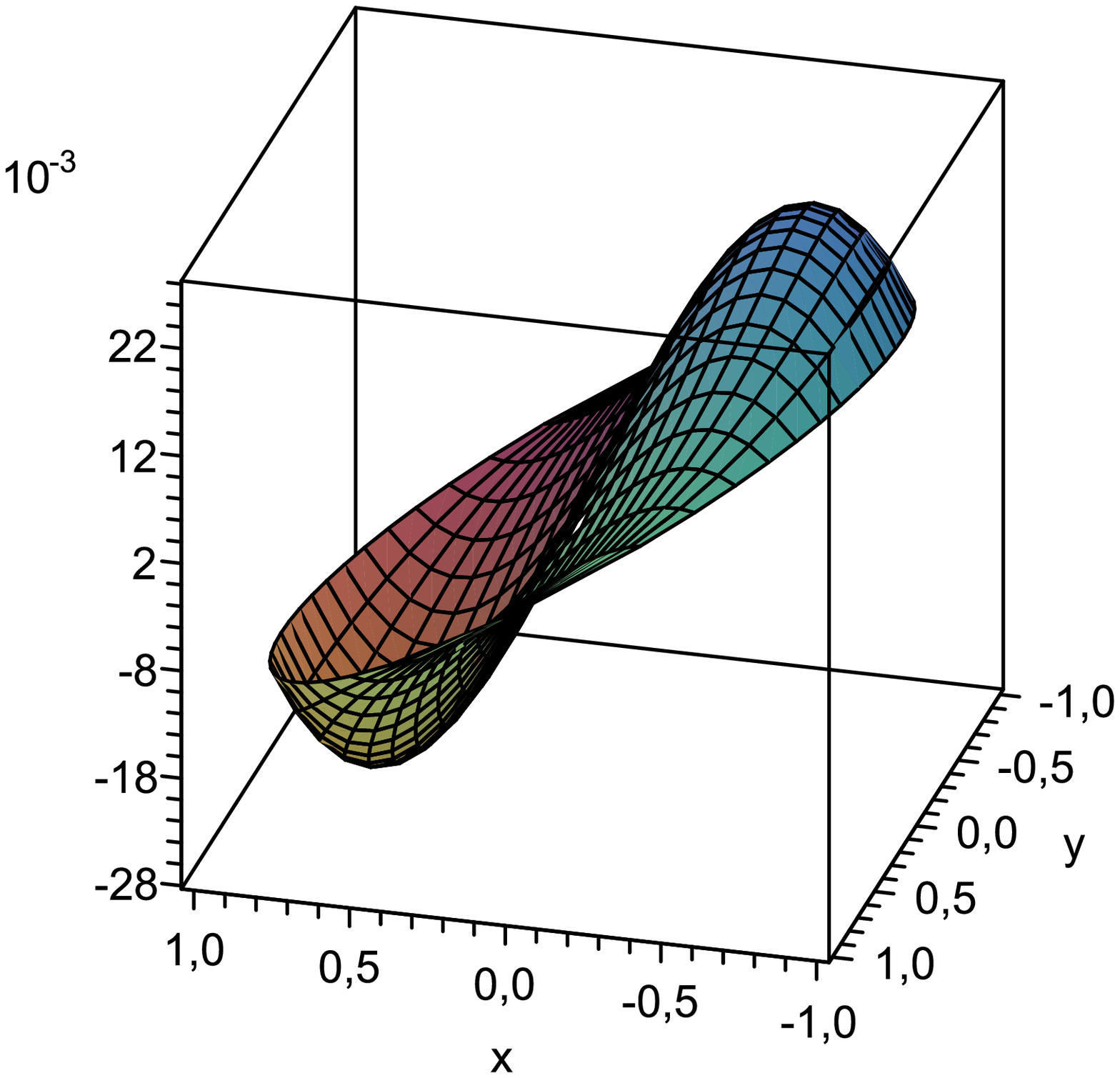}
\includegraphics[width=7 cm]{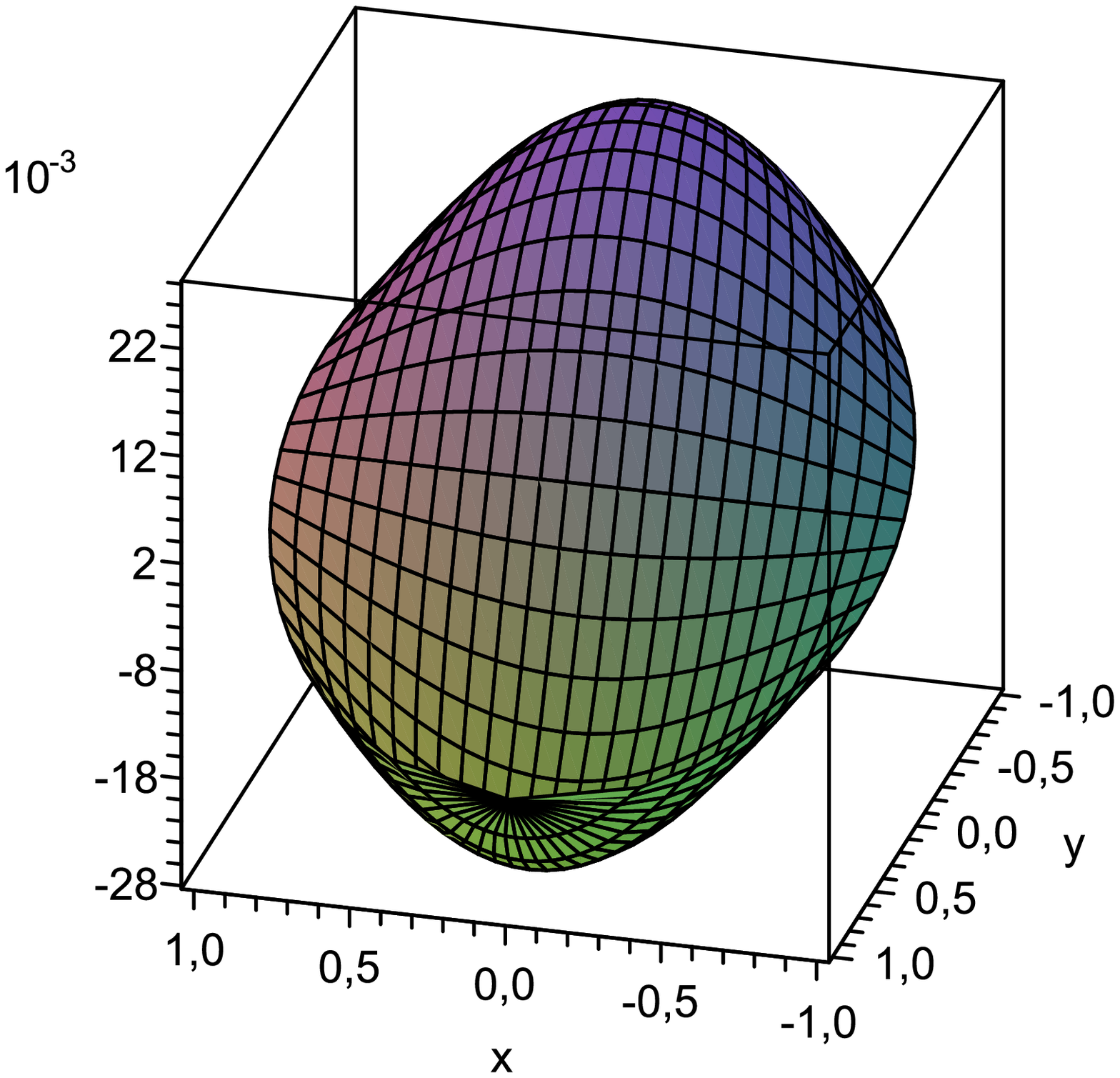}
\caption{ Surfaces representing two components  $u_1$ (left) and
$u_2$ (right) of the cell velocity in the time moment $t=2$ for  the
parameters $d_0 = 0.75, c_1=1, c_3 = 0.5, c_4 = 5, n = 3,
\sigma_0=-3, \delta=1$ and $ \lambda = 4$ (see (\ref{4-13})).}
\label{f1}
\end{figure}

\begin{figure}
\centering
\includegraphics[width=7 cm]{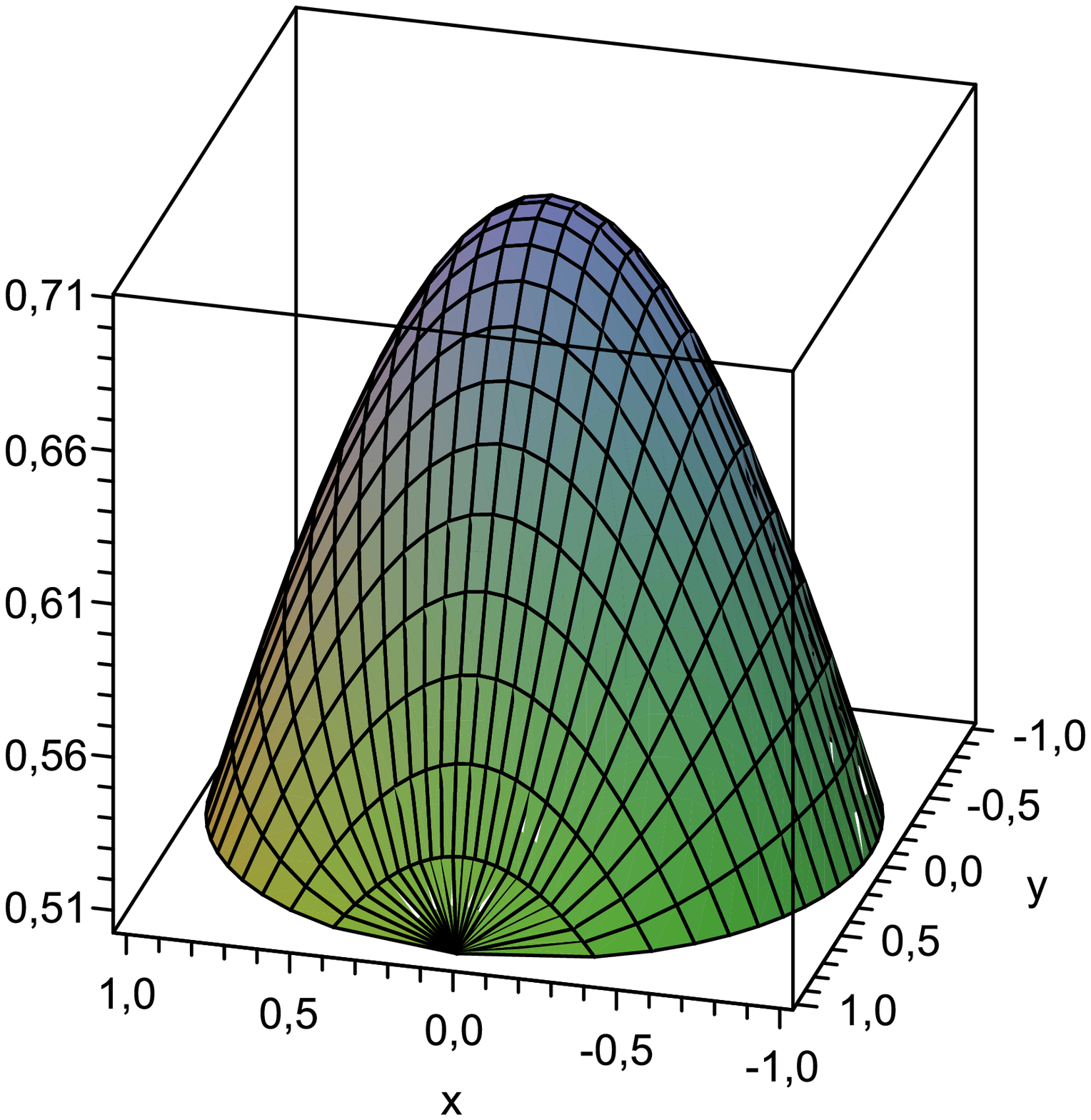}
\includegraphics[width=7 cm]{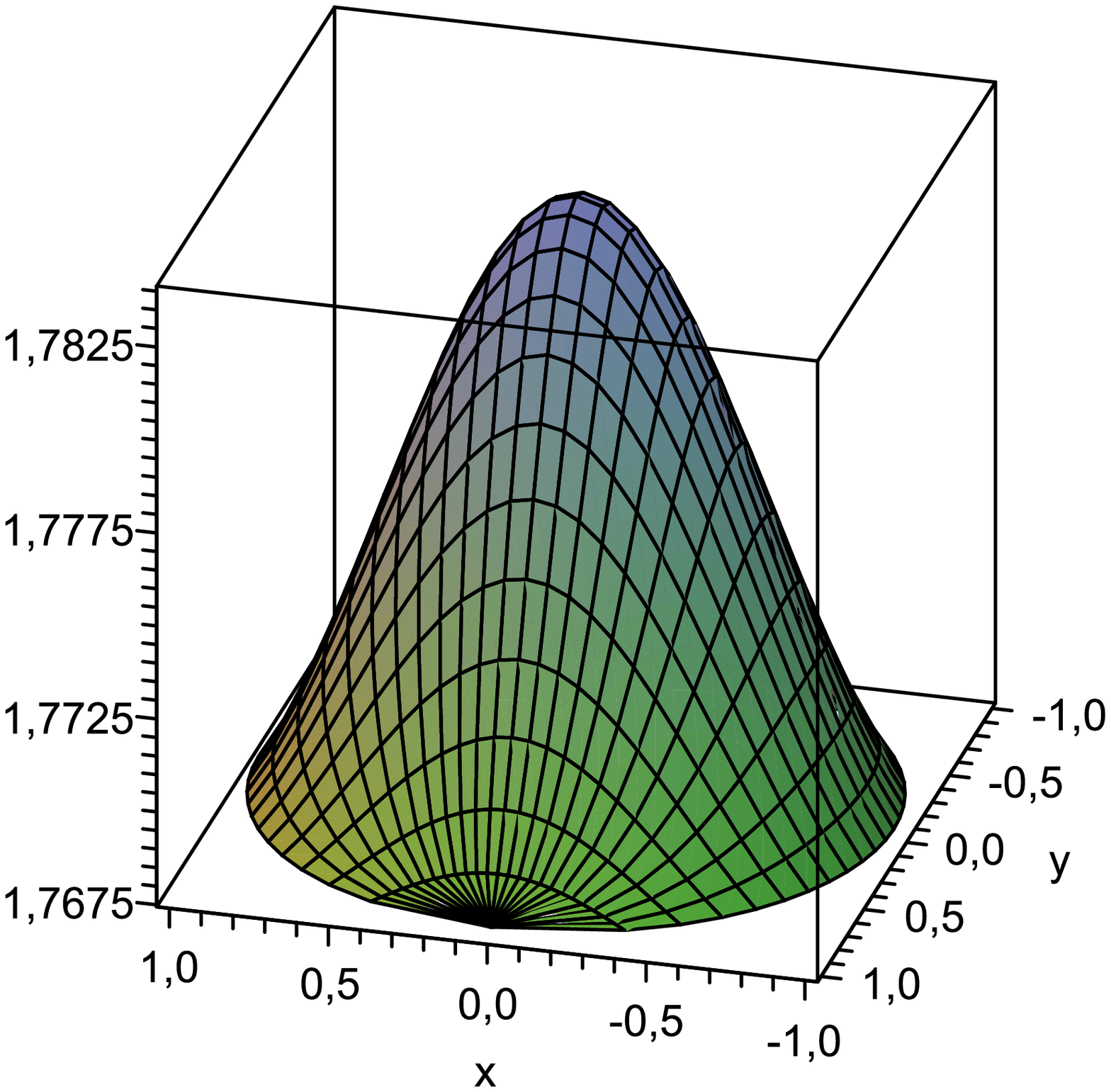}
\caption{ Surfaces representing the cell concentration   $\alpha$
(left) and the pressure $p$ (right) in the time moment $t=2$ for the
parameters $d_0 = 0.75, c_1=1, c_3 = 0.5, c_4 = 5, n = 3,
\sigma_0=-3, \delta=1$ and $ \lambda = 4$ (see (\ref{4-13})).}
\label{f2}
\end{figure}

 One notes that solution
 (\ref{4-13}) possesses a singularity in the point $(x,y)=(0,0)$.
 In order to avoid this singularity one needs to specify arbitrary
 parameter $c_3$ as follows:
 \be\label{4-38}c_3=\frac{2\sigma_0c_1^{n}}{(n-1)(2+
 \lambda)}+\frac{2c_1}{n-1}.\ee
 Formula  (\ref{4-38}) was derived by the Taylor  expansions of the
 exponents in the RHS of formulae~(\ref{4-13}).
An example of the 3D plots of the exact solution (\ref{4-13}) with
the  coefficients satisfying (\ref{4-38}) and a fixed time is
presented in Figures~\ref{f1} and \ref{f2}.

Now we turn to the  nonlinear   BVP (\ref{3-9}) and (\ref{3-10}).
Taking into account that $\Phi_*=2k\pi$ and $m=-1$ the boundary
conditions (\ref{3-10}) can be rewritten in the form
  \be\label{4-12} R_*=0, \ P_*=0, \ R_*'=0 \ee if $r=\delta.$ Solution (\ref{4-11})
   satisfies  conditions (\ref{4-12}) only under the condition
   $c_3\neq0$ (otherwise $R_*^2+ R_*'^2\neq0$). Hence the
  additional restrictions
\be\label{4-39}\delta=e^{-\frac{c_4}{c_3}}, \
c_1=\frac{nc_3}{2}\,\exp\lf(\frac{e^{-\frac{2c_4}{c_3}}}{4d_0}\rg)\equiv\frac{nc_3}{2}\,E,
\ \sigma_0=-\frac{(2+\lambda)c_3}{2}\,\lf(\frac{2}{nc_3}\rg)^n \ee
are needed in order to satisfy (\ref{4-12}). It should be noted that
compatibility of condition (\ref{4-13}) with these restrictions
cannot be derived for any fixed  $c_i$.

Thus, the exact solution of the nonlinear BVP (\ref{2-1*}),
(\ref{2-2}) with \be\label{4-40}m=-1, \
s_0=\frac{n\sigma_0}{(n-1)(2+\lambda)},
\sigma_0=-\frac{(2+\lambda)c_3}{2}\,\lf(\frac{2}{nc_3}\rg)^n, \
\delta=e^{-\frac{c_4}{c_3}}, \  c_3(n-1)\not=0\ee has the form
\be\label{4-13**}\ba
u^1=\frac{2d_0}{nE}\frac{x}{t(x^2+y^2)}\lf[\exp\lf(\frac{x^2+y^2}{4d_0}\rg)+
\frac{E^n}{n-1}\,\exp\lf(\frac{(1-n)(x^2+y^2)}{4d_0}\rg)-\frac{nE}{n-1}\rg],
\medskip
\\
u^2=\frac{2d_0}{nE}\frac{y}{t(x^2+y^2)}\lf[\exp\lf(\frac{x^2+y^2}{4d_0}\rg)+
\frac{E^n}{n-1}\,\exp\lf(\frac{(1-n)(x^2+y^2)}{4d_0}\rg)-\frac{nE}{n-1}\rg],
\medskip
\\
p=t^\frac{n}{1-n}\lf[\frac{c_3E^n}{1-n}\,\mbox{\raisebox{-3.2ex}{$\stackrel{\displaystyle\int^\delta}{\scriptstyle
\sqrt{x^2+y^2}\hskip0.2cm}$}}
 \frac{\exp\lf(-\frac{nz^2}{4d_0}\rg)}{z}\,dz+
\frac{c_3nE}{n-1}\mbox{\raisebox{-3.2ex}{$\stackrel{\displaystyle\int^\delta}{\scriptstyle
\sqrt{x^2+
y^2}\hskip0.2cm}$}}\frac{\exp\lf(-\frac{z^2}{4d_0}\rg)}{z}\,dz+c_4+\frac{c_3}{2}\ln (x^2+y^2)\rg],\\
\alpha=\frac{c_3nE}{2}\,t^\frac{1}{1-n}\exp\lf(-\frac{x^2+y^2}{4d_0}\rg), \\
\Gamma=x^2+y^2-\delta^2. \medskip
 \ea\ee

\begin{figure}
\centering
\includegraphics[width=7 cm]{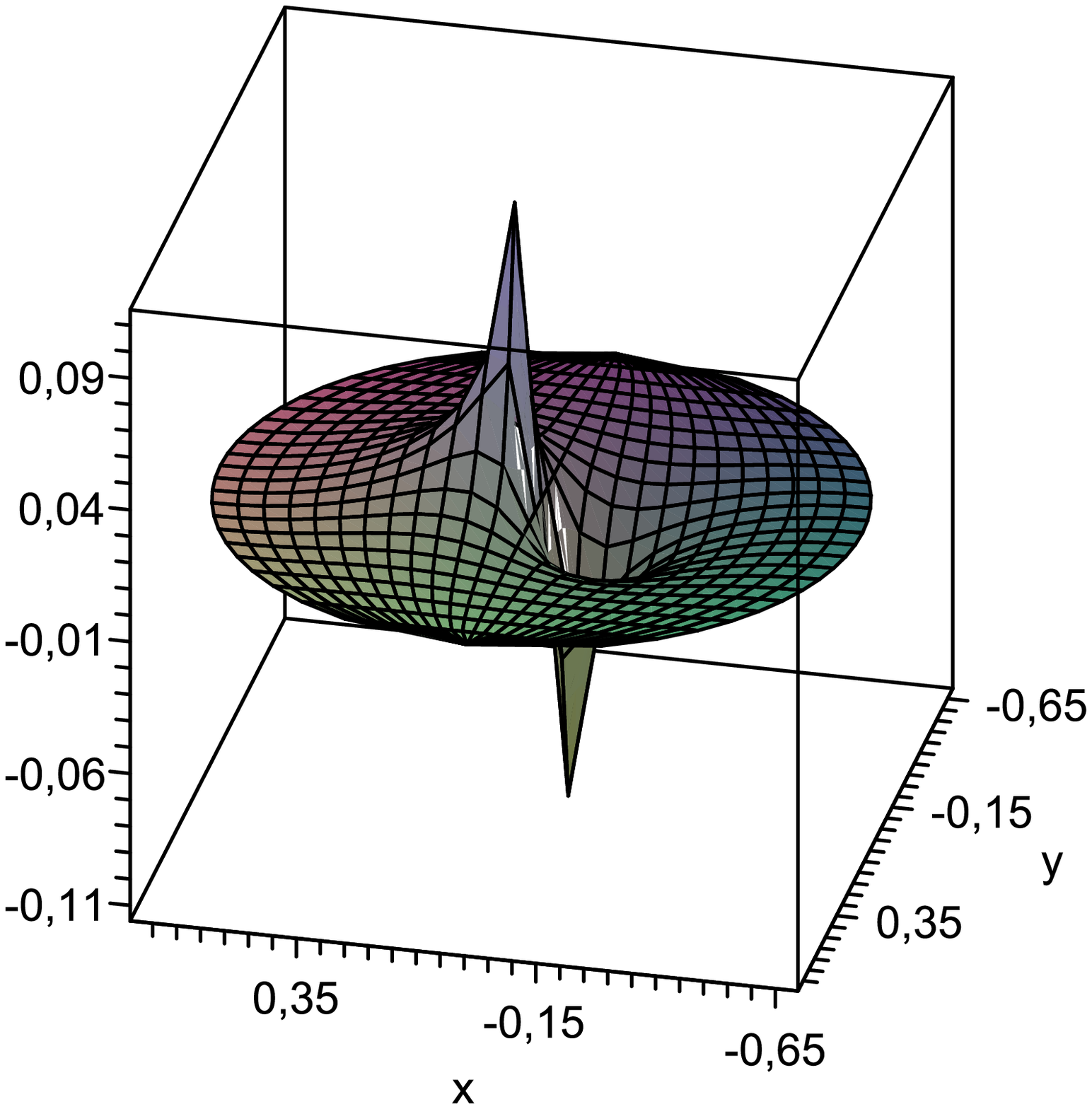}
\includegraphics[width=7 cm]{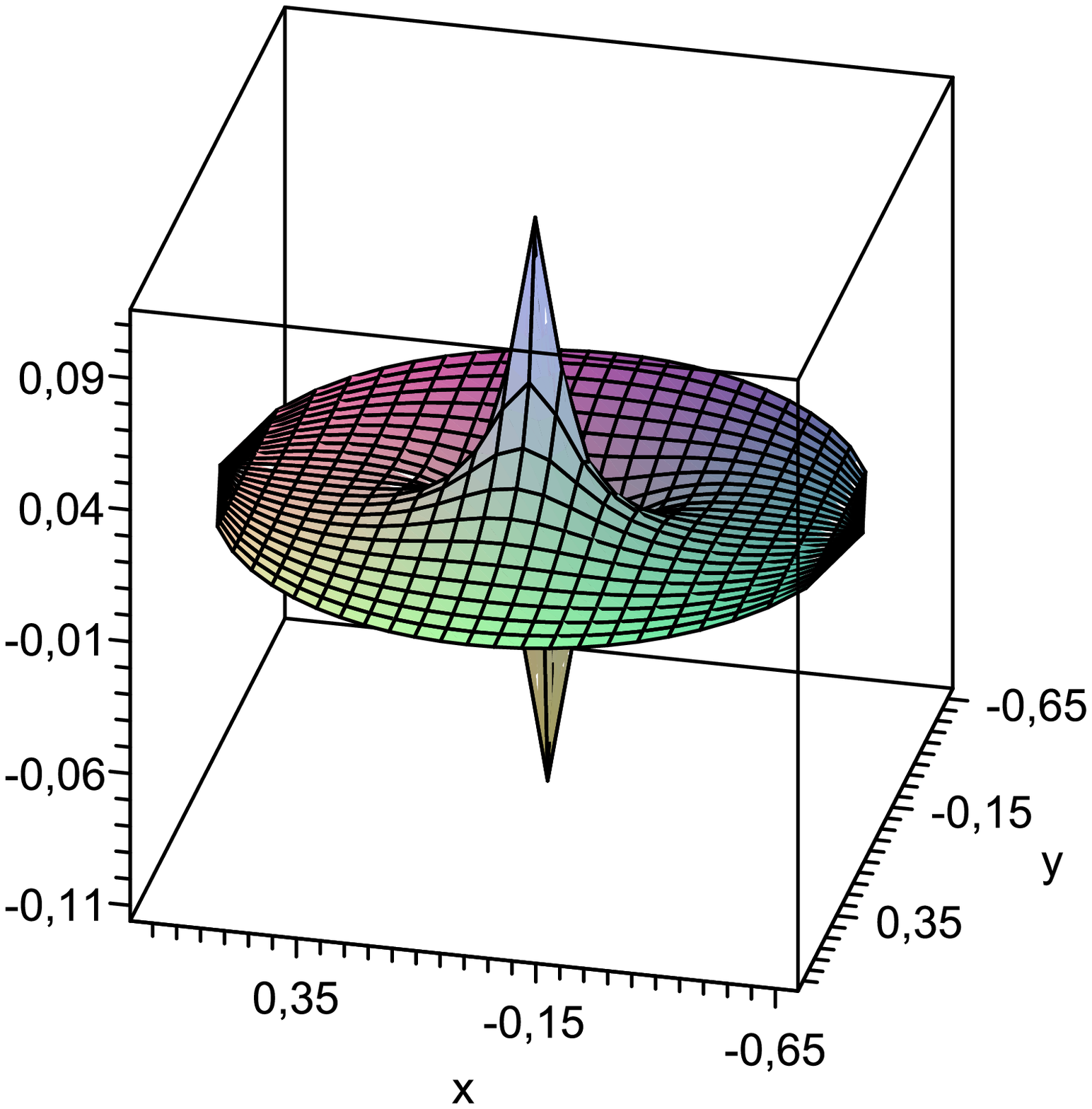}
\caption{ Surfaces representing two components  $u_1$ (left) and
$u_2$ (right) of the cell velocity in the time moment $t=1$ for  the
parameters $d_0 = 2, c_3 = 5, c_4 = 2, n = 2$ and $ \lambda = 4$
(see (\ref{4-13**})).} \label{f3}
\end{figure}

\begin{figure}
\centering
\includegraphics[width=7 cm]{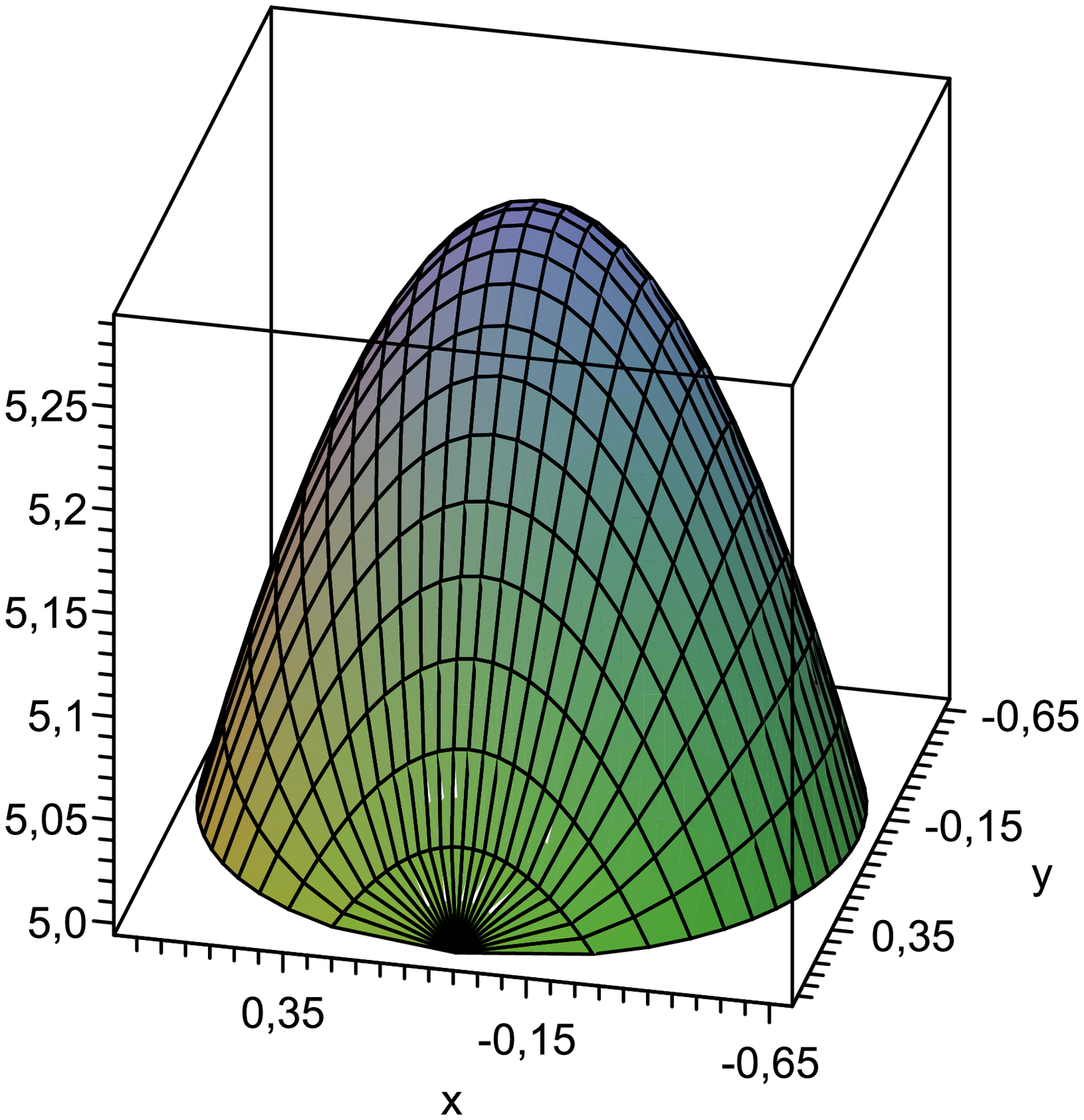}
\includegraphics[width=7 cm]{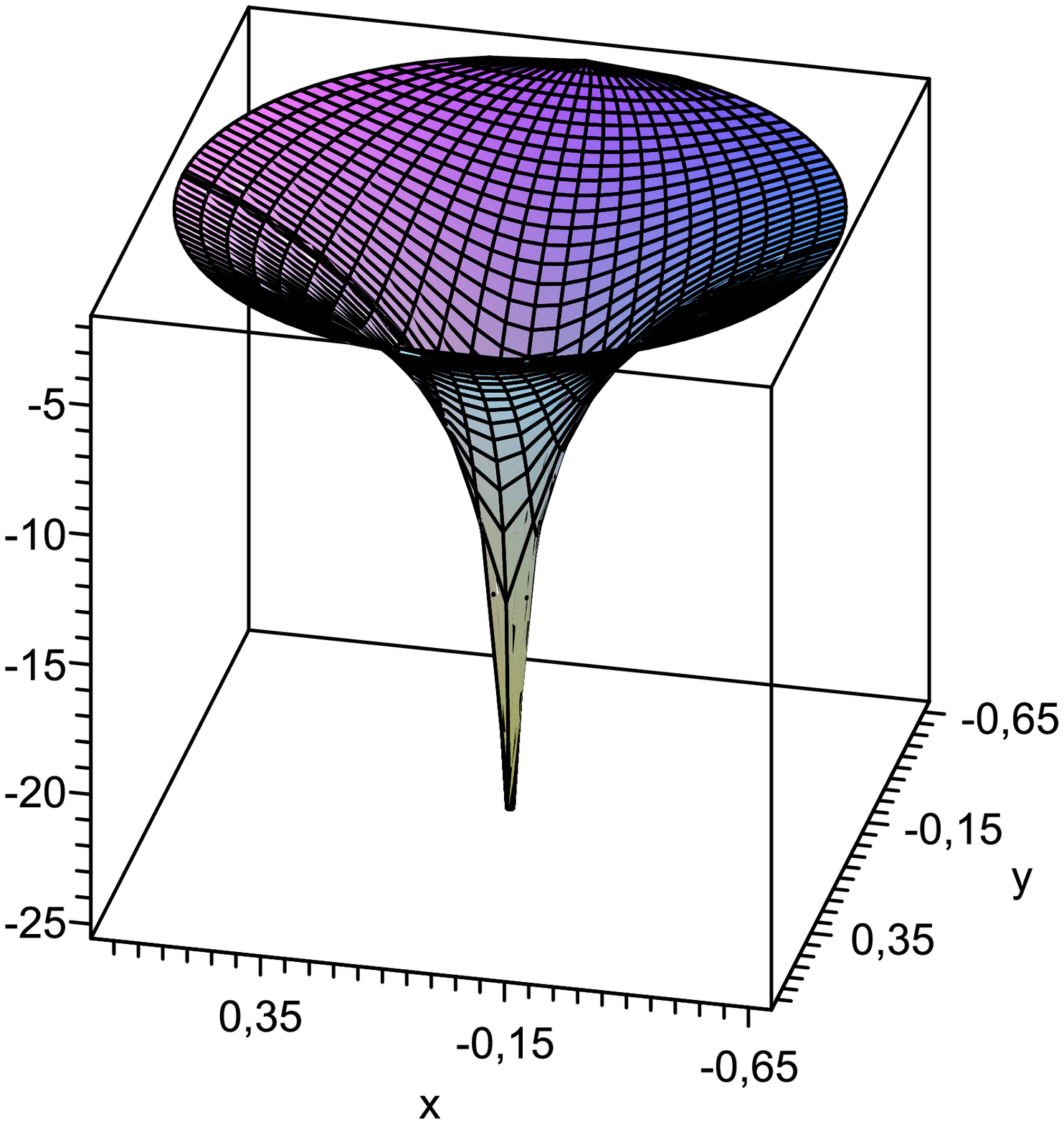}
\caption{ Surfaces representing the cell concentration   $\alpha$
(left) and the pressure $p$ (right) in the time moment $t=1$  for
the  parameters $d_0 = 2, c_3 = 5, c_4 = 2, n = 2, \lambda = 4$ and
$\delta\approx0.67$ (see (\ref{4-13**})).} \label{f4}
\end{figure}

\begin{figure}
\centering
\includegraphics[width=8 cm]{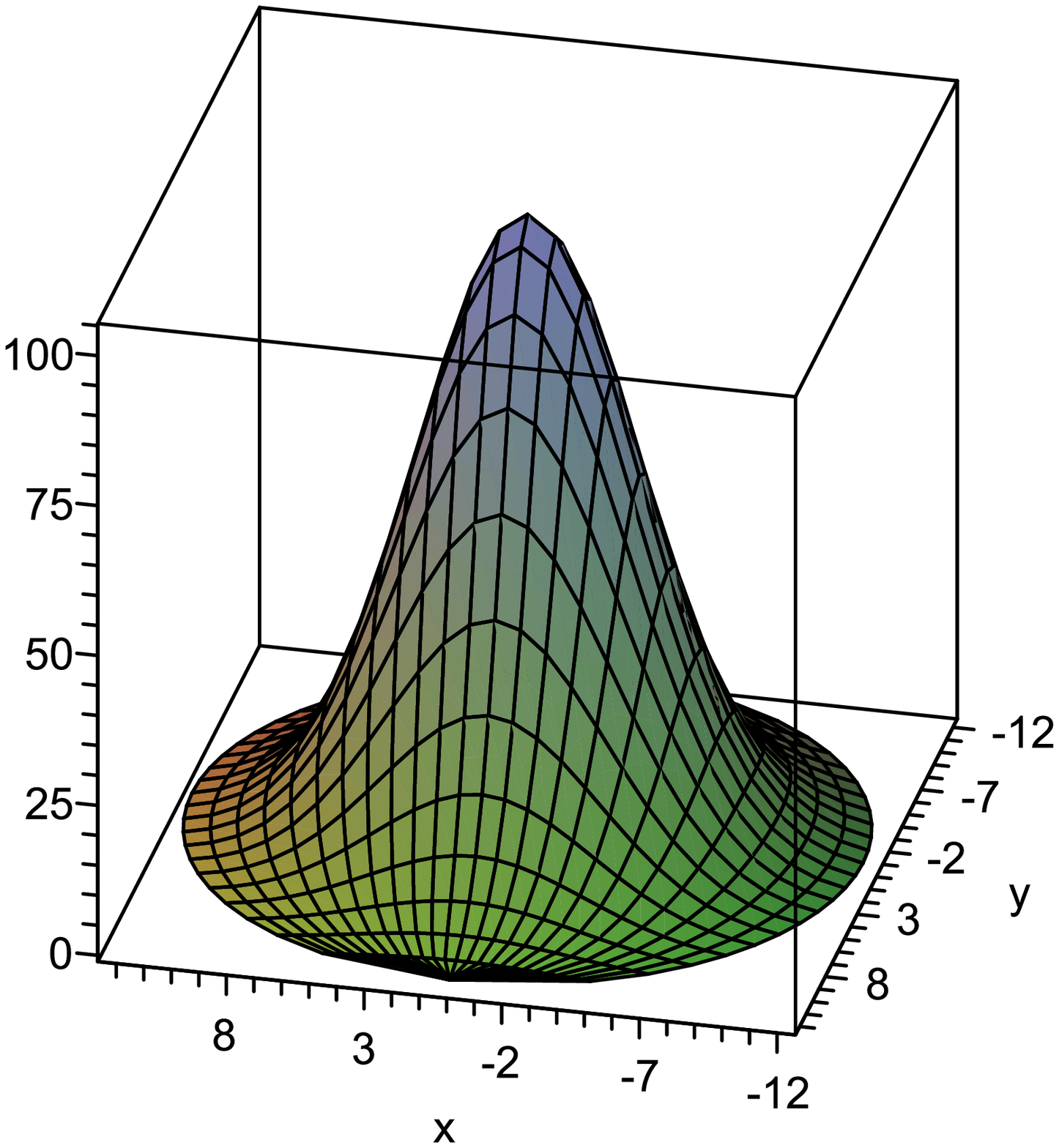}
\includegraphics[width=6 cm]{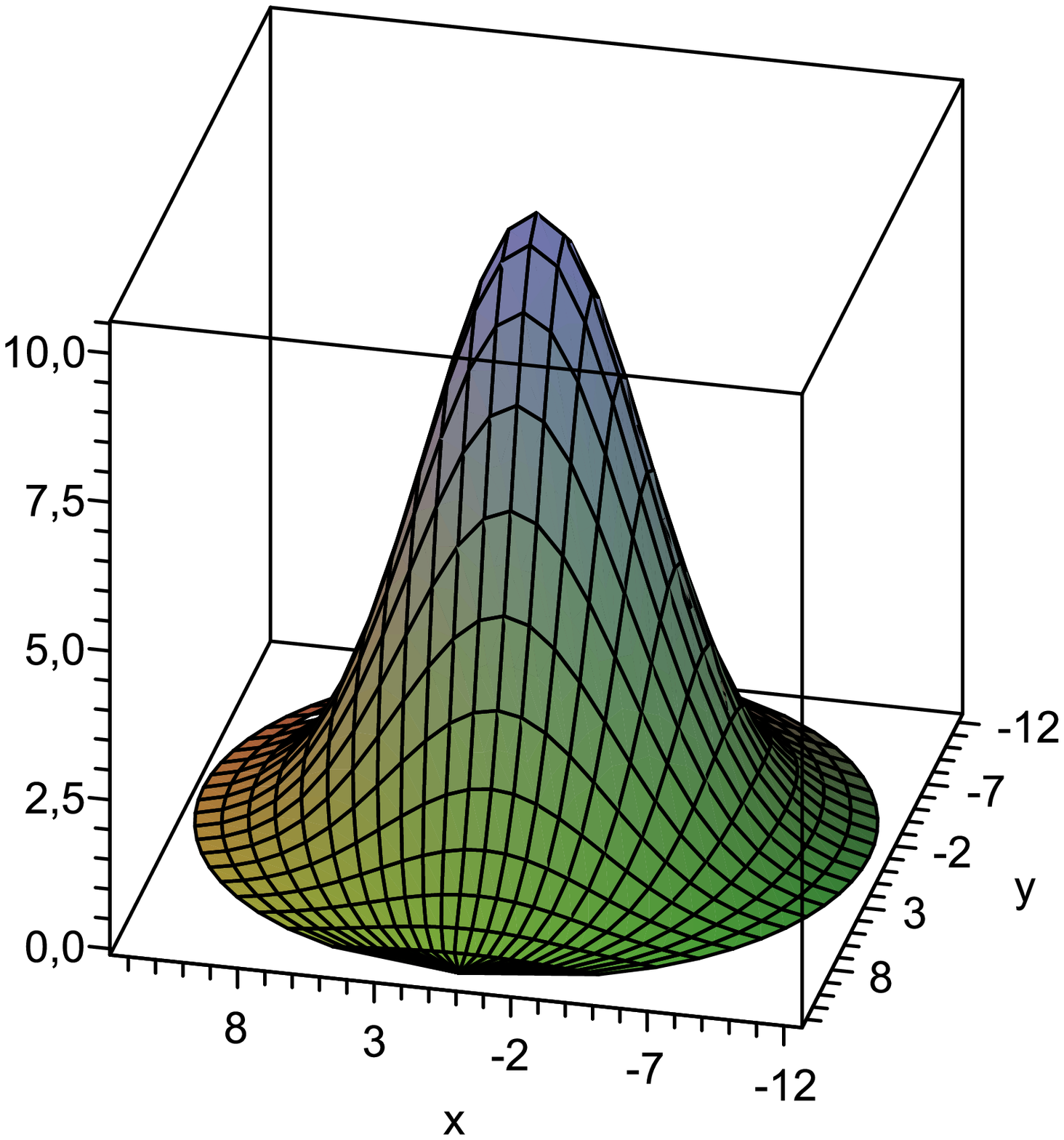}
\caption{ Surfaces representing the cell concentration   $\alpha$
 in the time moments $t=1$ (left) and  $t=10$ (right) for
the  parameters $d_0 = 8, c_3 = 1, c_4 = -2.5, n = 2, \lambda = 4$
and $\delta\approx12.18$ (see (\ref{4-13**})).} \label{f5}
\end{figure}

This solution with the parameters satisfying conditions (\ref{4-40})
is presented in Fig.\,\ref{f3} and Fig.\,\ref{f4}. It should be
stressed that the boundary $\Gamma$ is not moving in time, so that
one may interprets that the solution describes the solid tumour
growth at its final stage (no resources for further expansion). We
also note that the parameter $c_3n$ must be positive in order to
have the positive cell concentration
  $\alpha$ (see the fourth formula in (\ref{4-13**})).
Moreover, assuming that the  cell concentration  decreases with
time, we set $n>1$, hence $c_3 > 0, \ \sigma_0 < 0$  and $s_0 < 0$
(see conditions (\ref{4-40})). Thus, formulae (\ref{4-13**}) present
the exact solution  when the functions $\Sigma(\alpha)$ and
$S(\alpha)$ are negative. Obviously, the cell concentration
  $\alpha \to 0$  as $t \to \infty$ and this means that tumour is dying.  Notably, the  concentration plot  possesses  two different  forms depending on  the tumour radius $\delta$
  and time. The plots of the cell concentration presented in Fig.\,\ref{f4} and Fig.\,\ref{f5} show this difference in the forms.

\begin{remark}  The exact solution  (\ref{4-13**}) is highly nontrivial.
However the pressure $p$ and the cell velocity $(u_1, u_2)$ are
unbounded in the point $x=y=0$ because formulae (\ref{4-38}) and
(\ref{4-39}) cannot be satisfied by any choice of the parameters $c_i$. In other words,   restrictions (\ref{4-38}) and
(\ref{4-39}) are not compatible therefore this singularity cannot be avoided. From the physical point of view,
 it means that we deal with the water  flow, which can be approximated   by
 the  classical radially symmetric flow (see Remark~\ref{rem-3} above). Thus, such singularity is natural.
\end{remark}

We remind the reader that the exact solution (\ref{4-40}) was  derived  under  the
restriction $m=-1$ (see (\ref{4-7})). Let us consider the case
$m\neq-1.$ In this case, we start from formula (\ref{4-8}). Making
the similar examination as it was done above for the case $m=-1$,
new exact solutions of BVP (\ref{2-1*}), (\ref{2-2}) can be derived.
However, some coefficient restrictions are again needed. As a
result, two cases occur leading to two different exact solutions:

\be\label{4-42}\ba u^1=x\left(x^2+y^2\right)^{-\frac{2+m}{1+m}}
\left[d_0c_2c_1^mt^{\frac{1+m+n}{1-n}}+\frac{s_0(1+m)c_1^{n-1}}{2(1+m+n)}\left(x^2+y^2\right)^{\frac{1+m+n}{1+m}}\rg]
,\medskip
\\
u^2=y\left(x^2+y^2\right)^{-\frac{2+m}{1+m}}
\left[d_0c_2c_1^mt^{\frac{1+m+n}{1-n}}+\frac{s_0(1+m)c_1^{n-1}}{2(1+m+n)}\left(x^2+y^2\right)^{\frac{1+m+n}{1+m}}\rg],\medskip
\\
p=\frac{ s_0(1+m)^2
c_1^{n-1-m}}{4d_0n(1+m+n)}\,\left(x^2+y^2\right)^{\frac{n}{1+m}}-\frac{c_2}{2}\,\frac{t^{\frac{1+m+n}{1-n}}
}{ x^2+y^2}+c_3t^{\frac{n}{1-n}} , \medskip\\
\alpha=c_1(x^2+y^2)^{\frac{1}{1+m}}, \
\Gamma=x^2+y^2-\delta^2t^{\frac{1+m}{1-n}}, \ea\ee if
\be\nonumber\ba m\neq -n-1,\ d_0=\frac{1+m}{4 (1+\lambda
)}\,c_1^{-1-m},\ s_0=\frac{n\sigma_0}{(n-1) (2+\lambda )}, \
\sigma_0=-\frac{c_1^{1-n}  (3+m+\lambda )}{n}\,\delta ^{\frac{2-2
n}{1+m}},
 \medskip\\
c_2=\frac{2c_1 (1+\lambda ) \lf[-1+m+2 n+\lambda(m+n)  \rg]}{(1-n)
(1+m+n) (2+\lambda )}\,\delta ^{2+\frac{2}{1+m}}, \ c_3=\frac{c_1
 (1+\lambda ) \lf[3+m+\lambda -n (2+\lambda
)\rg]}{n(n-1)(2+\lambda )}\,\delta ^{\frac{2}{1+m}}, \
n(n-1)\neq0,\ea\ee
and \be\label{4-44}\ba u^1=\frac{x
\left(x^2+y^2\right)^\frac{1-n}{n}}{2c_1^{1+n}}\,\lf[2d_0c_2+s_0c_1^{2n}\ln\lf(t^{\frac{n}{1-n}}(x^2+y^2)\rg)\rg],\\
u^2=\frac{y
\left(x^2+y^2\right)^\frac{1-n}{n}}{2c_1^{1+n}}\,\lf[2d_0c_2+s_0c_1^{2n}\ln\lf(t^{\frac{n}{1-n}}(x^2+y^2)\rg)\rg],\medskip
\\
p=-\frac{1}{4d_0(x^2+y^2)}\,\lf[2d_0c_2+s_0c_1^{2n}\lf[1+\ln\lf(t^{\frac{n}{1-n}}(x^2+y^2)\rg)\rg]\rg]+c_3t^{\frac{n}{1-n}},
\medskip\\ \alpha=c_1(x^2+y^2)^{-\frac{1}{n}}, \
\Gamma=x^2+y^2-\delta^2t^{\frac{n}{n-1}}, \medskip
 \ea\ee if \be\nonumber\ba m=-n-1, \  d_0=-\frac{nc_1^n}{4 (1+\lambda)},\ s_0=\frac{n\sigma_0}{(n-1) (2+\lambda )},
  \ \sigma_0=\frac{(n-2-\lambda ) }{n}\,c_1^{1-n} \delta ^{2-\frac{2}{n}}
,   \medskip\\
c_2=\frac{2c_1  (1+\lambda ) \lf[n (2+\lambda )+2 (2-n+\lambda )
\ln\delta\rg]}{n(1-n)(2+\lambda )}\,\delta ^{2-\frac{2}{n}}, \
c_3=\frac{c_1  (1+\lambda ) \lf[2+\lambda -n (3+\lambda
)\rg]}{n(n-1)(2+\lambda )}\,\delta ^{-\frac{2}{n}}, \
n(n-1)\neq0.\ea\ee

 In contrast to the exact solution (\ref{4-13**}), the exact solutions (\ref{4-42}) and (\ref{4-44}) involve the  boundary $\Gamma$ moving in time, however, the cell concentration $\alpha$ does not depend on time. It can be easily seen that  a singularity again occurs in the  point $(x,y)=(0,0)$.

\section{\bf The general case} \label{s5}

In this section we present a preliminary Lie symmetry analysis of
 the nonlinear BVP (\ref{2-1})--(\ref{2-2}) in the general case, i.e. restrictions (\ref{2-5}) are not applied in what follows.
Let us apply   the Lie symmetry operator \be\label{4-13*} J =
x\partial_y -y\partial_x + u^1\partial_{u^2}- u^2\partial_{u^1}\ee
for reducing the problem to the lower dimensionality.

  First of all, we rewrite
the nonlinear BVP (\ref{2-1})--(\ref{2-2}) in the polar coordinates
applying the formulae
 \be\label{4-14}\begin{array}{l} x=r\cos\phi, \  y=r\sin\phi,\\
 u^1=R(t,r,\phi)\cos\Phi(t,r,\phi),\ u^2=R(t,r,\phi)\sin\Phi(t,r,\phi),\\
 \alpha=\Lambda(t,r,\phi), \ p=P(t,r,\phi)
 \end{array}\ee  in order to
simplify  further computations. Obviously, formulae (\ref{4-14})
transforms operator (\ref{4-13*}) to the form
\be\label{4-15}J=-\p_{\phi}-\p_{\Phi}.\ee The ansatz corresponding
to operator (\ref{4-15}) can be easily derived \be\label{4-16}\ba
R=R^*(t,r), \ \Phi=\Phi^*(t,r)+\phi, \ \Lambda=\Lambda^*(t,r), \
P=P^*(t,r), \ea\ee where the functions with upper stars are new
unknown those. Thus, substituting (\ref{4-14}) and (\ref{4-16}) into
(\ref{2-1}), we obtain
the two-dimensional governing equations
 \be\label{4-17}\ba r\Lambda^*_t+\Big(r\Lambda^*R^*\cos\Phi^*\Big)_r=rS(\Lambda^*),\medskip\\
\Big(rR^*\cos\Phi^*\Big)_r=\Big(rD(\Lambda^*)P^*_r\Big)_r,\medskip\\
(1+\lambda)R^* \Lambda^*_r\sin2\Phi^*-(2+\lambda)\Big(rR^*
\Lambda^*\Phi^*_r\Big)_r-(2+\lambda)r\Lambda^*R^*_r\Phi^*_r=\\
\hskip3cm
r\Big(\frac{d}{d\Lambda^*}\lf(\Lambda^*\Sigma(\Lambda^*)\rg)+
P^*_r\Big)\sin\Phi^*,\medskip\\
(1+\lambda)rR^*
\Lambda^*_r\cos2\Phi^*+(2+\lambda)r\Big(r\Lambda^*R^*_r\Big)_r-(2+\lambda)
\Lambda^*R^*\Big(1+r^2{\Phi^*_r}^2\Big)-rR^*
\Lambda^*_r =\\
\hskip3cm
r^2\Big(\frac{d}{d\Lambda^*}\lf(\Lambda^*\Sigma(\Lambda^*)\rg)+P^*_r\Big)\cos\Phi^*.\ea\ee

In order to reduce  the boundary conditions (\ref{2-2}), one firstly needs to
specify the function $\Gamma$. Rewriting  $\Gamma$ in the polar
coordinates, one easily checks that one is invariant under operator
(\ref{4-15}) if  \be\label{4-18}\Gamma\equiv
\Gamma^*(t,r)=0.\ee
So, using (\ref{4-14}), (\ref{4-16}) and (\ref{4-18}),    the boundary conditions (\ref{2-2}) are reduced to
  \be\label{4-19}\ba \Gamma^*_t+R^*\Gamma^*_r\cos\Phi^*=0, \ P^*=0,\medskip\\
(2+\lambda)rR^*_r+R^*\Big((1+\lambda)\cos2\Phi^*-1\Big)=0,\medskip\\
R^*\Big((2+\lambda)r\Phi^*_r-(1+\lambda)\sin2\Phi^*\Big)=0.\ea\ee

The two-dimensional BVP (\ref{4-17}), (\ref{4-19}) is still a
nonlinear problem with the moving boundary and its exact solving is
a highly  complicated task. Here we restrict ourselves to search for
stationary (i.e. steady-state) solutions. From the point of view of
Lie method, it means application of   the time translation  operator
$\partial_t$ for reducing BVP (\ref{4-17}), (\ref{4-19}) to ODE
problem.  The ansatz corresponding to this operator is
\be\label{4-33}\ba R^*(t,r)=R_*(r), \ \Phi^*(t,r)= \Phi_*(r), \
\Lambda^*(t,r)=\Lambda_*(r), \ P^*(t,r)=P_*(r), \ea\ee where the
functions with the lower stars are new unknown those. Thus, we
obtain the following BVP with the governing ODEs
\be\label{4-34}\ba \Big(r\Lambda_*R_*\cos\Phi_*\Big)'=rS(\Lambda_*),\medskip\\
\Big(rR_*\cos\Phi_*\Big)'=\Big(rD(\Lambda_*)P_*'\Big)',\medskip\\
(1+\lambda)R_* \Lambda_*'\sin2\Phi_*-(2+\lambda)\Big(rR_*
\Lambda_*\Phi_*'\Big)'-(2+\lambda)r\Lambda_*R_*'\Phi_*'=\\
\hskip3cm
r\Big(\frac{d}{d\Lambda_*}\lf(\Lambda_*\Sigma(\Lambda_*)\rg)+
P_*'\Big)\sin\Phi_*,\medskip\\
(1+\lambda)rR_*
\Lambda_*'\cos2\Phi_*+(2+\lambda)r\Big(r\Lambda_*R_*'\Big)'-(2+\lambda)
\Lambda_*R_*\Big(1+r^2{\Phi_*'}^2\Big)-rR_*
\Lambda_*' =\\
\hskip3cm
r^2\Big(\frac{d}{d\Lambda_*}\lf(\Lambda_*\Sigma(\Lambda_*)\rg)+P_*'\Big)\cos\Phi_*,\ea\ee
(here the upper prime means differentiation w.r.t. the variable $r$)
and the boundary conditions
\be\label{4-35}\ba r=\delta: \quad R_*\cos\Phi_*=0, \ P_*=0,\medskip\\
r=\delta: \quad (2+\lambda)rR_*'+R_*\Big((1+\lambda)\cos2\Phi_*-1\Big)=0,\medskip\\
r=\delta: \quad
R_*\Big((2+\lambda)r\Phi_*'-(1+\lambda)\sin2\Phi_*\Big)=0.\ea\ee

Now we again use  the additional restriction  (\ref{4-0}), i.e. $\sin\Phi_*=0$, hence  the
function \be\label{4-21}R_*=\frac{\beta}{r}+D(\Lambda_*)P_*'\ee
(here $\beta$ is an arbitrary constant) immediately follows  from the second equation of
system (\ref{4-34}). Other equations of (\ref{4-34}) take the form
(the third equation vanishes)
\be\label{4-22}\ba \Big(r\Lambda_*R_*\Big)'=rS(\Lambda_*),\medskip\\
\lambda rR_*
\Lambda_*'+(2+\lambda)r\Big(r\Lambda_*R_*'\Big)'-(2+\lambda)
\Lambda_*R_*=
r^2\Big(\frac{d}{d\Lambda^*}\lf(\Lambda^*\Sigma(\Lambda^*)\rg)+P_*'\Big).\ea\ee

Substituting (\ref{4-21}) into system (\ref{4-22}), we obtain the
nonlinear ODE system with respect to the functions $\Lambda_*$ and
$P_*$. Using the analogous procedure as in previous section, one can
construct the overdetermined system \be\label{4-23}\ba
\Lambda_*''-\Lambda_*^{-1}{\Lambda_*'}^2-
\frac{\lambda}{(2+\lambda)r}\,\Lambda_*'+\frac{1}{(2+\lambda)D(\Lambda_*)}=0,\\
D(\Lambda_*)\lf(\frac{S(\Lambda_*)}{\Lambda_*}-\frac{dS(\Lambda_*)}{d\Lambda^*}
+\frac{1}{2+\lambda}\frac{d}{d\Lambda^*}\lf(\Lambda^*\Sigma(\Lambda^*)\rg)\rg)\Lambda_*'
=\frac{\beta}{(2+\lambda)r}.\ea\ee

System (\ref{4-23}) consist of two nonlinear equations for finding
the function $\Lambda_*$. To solve this system, one needs to specify
the functions $D(\Lambda_*), \ S(\Lambda_*)$ and $\Sigma(\Lambda_*)$
otherwise one is not integrable. We aim to find nontrivial
steady-state solutions of BVP (\ref{2-1})--(\ref{2-2}). In order to
construct them in explicit form, we set
\be\label{4-24}D(\Lambda_*)=d_0\Lambda_*^{-1}. \ee In this case, the
general solution of the first equation of system (\ref{4-23}) has
the form
\be\label{4-25}\Lambda_*=c_1\exp\lf(c_2r^{\frac{2+2\lambda}{2+\lambda}}-\frac{r^2}{4d_0}\rg),\ee
where $c_1>0$ (because $\Lambda_*$ means the cell density) and $c_2$
are arbitrary constants. Substituting (\ref{4-25}) into the second
equation of system (\ref{4-23}),  one
 obtains the functional-differential equation
\be\nonumber\frac{S(\Lambda_*)}{\Lambda_*}-\frac{dS(\Lambda_*)}{d\Lambda_*}
+\frac{1}{2+\lambda}\frac{d}{d\Lambda^*}\lf(\Lambda_*\Sigma(\Lambda_*)\rg)=
\frac{2\beta}{4d_0c_2(1+\lambda)r^{\frac{2+2\lambda}{2+\lambda}}-(2+\lambda)r^2}.\ee
 Because the functions $S$ and $\Sigma$ does not depend explicitly on the variable $r$,  this equation has solutions only under the restriction $\beta=0$.
So, we arrive at the linear ODE w.r.t. either $S$ and~$\Sigma$:
\be\label{4-27}\frac{S(\Lambda_*)}{\Lambda_*}-\frac{dS(\Lambda_*)}{d\Lambda_*}
+\frac{1}{2+\lambda}\frac{d}{d\Lambda_*}\lf(\Lambda_*\Sigma(\Lambda_*)\rg)=
0,\ee which can be easily solved.

Thus, the following solution of the nonlinear system (\ref{4-34}) with the  triplet $(S, \  D,\ \Sigma)$ satisfying restrictions (\ref{4-24})  and (\ref{4-27})
is derived: \be\label{4-37}\ba
\Phi_*=2k\pi, \ \Lambda_*=c_1\exp\lf(c_2r^{\frac{2+2\lambda}{2+\lambda}}-\frac{r^2}{4d_0}\rg),\\
P_*=c_4+c_3\ln r+\frac{1}{d_0}\int\lf(\frac{1}{r}\int rS(\Lambda_*)dr\rg)dr,\medskip\\
 R_*=\frac{d_0}{\Lambda_*}\lf(\frac{1}{d_0r}\int rS(\Lambda_*)dr+\frac{c_3}{r}\rg).\ea\ee

The  boundary conditions (\ref{4-35}) with the restriction (\ref{4-0}) are essentially simplified  and take the form
\be\label{4-35*} r=\delta: \  P_*=0, \  R_*=0, \ R_*'=0. \ee
 Obviously, the exact   solution (\ref{4-37}) satisfies boundary conditions (\ref{4-35*})
 provided the function $S$ (or $\Sigma$) is given and
  the constants $c_i$ are correctly-specified.

\medskip

\textbf{Example.} Let us set
\be\label{4-28}S(\Lambda_*)=k_1\Lambda_*^m-k_2\Lambda_*^n,\ee where
$k_1$ and $k_2>0$ are arbitrary constants,  $0<m<n$. Notably, the
above  function  for the net cell proliferation rate in the case
$m=1, \ n=2$ gives exactly the profile suggested in
\cite{by-ki-2003}. Then the function $\Sigma(\Lambda^*)$ has the
form (see equation (\ref{4-27}))
\be\label{4-29}\Sigma(\Lambda_*)=(2+\lambda)\lf[k_1\lf(1-\frac{1}{m}\rg)\Lambda_*^{m-1}+
k_2\lf(\frac{1}{n}-1\rg)\Lambda_*^{n-1}\rg].\ee

Substituting (\ref{4-21}), (\ref{4-24}) and (\ref{4-28}) into  the
first equation of system (\ref{4-22}) and setting
 $c_2=0$ (in order to
simplify the solution obtained), the function \be\nonumber\ba
P_*=c_4+c_3\ln r+\frac{2k_1c_1^m}{m}\,\int^\delta_r
\frac{\exp\lf(-\frac{mz^2}{4d_0}\rg)}{z}\,dz-\frac{2k_2c_1^n}{n}\,\int^\delta_r
\frac{\exp\lf(-\frac{nz^2}{4d_0}\rg)}{z}\,dz\ea\ee  can be easily
derived. Here $c_3$ and $c_4$ are arbitrary constants,  $r<\delta$
and the constant $\delta$ should be specified using the boundary
conditions (\ref{4-35}).

Thus, the solution of  system (\ref{4-34}) with the functions
$D(\Lambda_*), \ S(\Lambda_*)$ and $\Sigma(\Lambda_*)$  of the form
(\ref{4-24}),  (\ref{4-28}) and  (\ref{4-29}) is
\be\label{4-31}\ba
\Phi_*=2k\pi, \ \Lambda_*=c_1\exp\lf(-\frac{r^2}{4d_0}\rg),\\
P_*=c_4+c_3\ln r+\frac{2k_1c_1^m}{m}\,\int^\delta_r
\frac{\exp\lf(-\frac{mz^2}{4d_0}\rg)}{z}\,dz-\frac{2k_2c_1^n}{n}\,\int^\delta_r
\frac{\exp\lf(-\frac{nz^2}{4d_0}\rg)}{z}\,dz,\medskip\\
 R_*=\frac{d_0}{c_1r}\lf[c_3\exp\lf(\frac{r^2}{4d_0}\rg)-
\frac{2k_1c_1^{m}}{m}\,\exp\lf(\frac{(1-m)r^2}{4d_0}\rg)+
\frac{2k_2c_1^{n}}{n}\,\exp\lf(\frac{(1-n)r^2}{4d_0}\rg)\rg].\ea\ee


Solution (\ref{4-31}) satisfies  the boundary conditions
(\ref{4-35*}) if the parameters are specified as follows
\be\label{4-36} c_4=- c_3\ln \delta, \
 k_1=\frac{c_3mn}{2c_1^m(n-m)}\,\exp\lf(m\frac{\delta^2}{4d_0}\rg), \
 k_2=\frac{c_3mn}{2c_1^n(n-m)}\,\exp\lf(n\frac{\delta^2}{4d_0}\rg).\ee

Thus, using (\ref{4-14}),  (\ref{4-16}),  (\ref{4-33})  and
(\ref{4-31})  the exact  solution \be\label{4-32}\ba
u^1=\frac{d_0x}{c_1\lf(x^2+y^2\rg)}\lf[c_3\exp\lf(\frac{x^2+y^2}{4d_0}\rg)-
\frac{2k_1c_1^{m}}{m}\,\exp\lf(\frac{(1-m)(x^2+y^2)}{4d_0}\rg)+
\frac{2k_2c_1^{n}}{n}\,\exp\lf(\frac{(1-n)(x^2+y^2)}{4d_0}\rg)\rg],
\medskip
\\ u^2=\frac{d_0y}{c_1\lf(x^2+y^2\rg)}\lf[c_3\exp\lf(\frac{x^2+y^2}{4d_0}\rg)-
\frac{2k_1c_1^{m}}{m}\,\exp\lf(\frac{(1-m)(x^2+y^2)}{4d_0}\rg)+
\frac{2k_2c_1^{n}}{n}\,\exp\lf(\frac{(1-n)(x^2+y^2)}{4d_0}\rg)\rg],
\medskip
\\
p=c_4+\frac{c_3}{2}\ln (x^2+y^2)+
\frac{2k_1c_1^{m-1}}{m}\,\mbox{\raisebox{-3.2ex}{$\stackrel{\displaystyle  
\int^\delta}{\scriptstyle \sqrt{x^2+y^2}\hskip0.2cm}$}}
 \frac{\exp\lf(-\frac{mz^2}{4d_0}\rg)}{z}\,dz-
 \frac{2k_2c_1^{n-1}}{n}\,\mbox{\raisebox{-3.2ex}{$\stackrel{\displaystyle  
\int^\delta}{\scriptstyle \sqrt{x^2+y^2}\hskip0.2cm}$}}
 \frac{\exp\lf(-\frac{nz^2}{4d_0}\rg)}{z}\,dz,\\
\alpha=c_1\exp\lf(-\frac{x^2+y^2}{4d_0}\rg)
 \ea\ee
 (here  coefficient restrictions (\ref{4-36}) take place)
of BVP  (\ref{2-1})--(\ref{2-2}) with the functions $D(\Lambda_*), \
S(\Lambda_*)$ and $\Sigma(\Lambda_*)$  of the form (\ref{4-24}),
(\ref{4-28}) and  (\ref{4-29}), respectively,  has been constructed.
  Because it is  the steady-state solution, the boundary   $\Gamma(t,x,y)=x^2+y^2-\delta^2$ does not depend on time.

The steady-state solution (\ref{4-32})    of BVP
(\ref{2-1})--(\ref{2-2}) is highly nontrivial. However, this
solution possesses singularity in the point $(x,y)=(0,0)$ because
this singularity  cannot be removed  by any choice of the parameters
$m, n, c_3$ and $c_4$ provided they satisfy restrictions
(\ref{4-36}). So, we have the similar situation to that occurring in
Section~\ref{s4}. Notably, the exact solution (\ref{4-32}) is an
analog of (\ref{4-13**}). In fact, solution (\ref{4-13**}) with a
fixed time $t=t_0>0$ has the same structure as (\ref{4-32}),
although they solve the nonlinear BVP (\ref{2-1})--(\ref{2-2}) with
different functions $ S $ and $\Sigma$. From the physical
point of view, it   means that a generalization of the   radially
symmetric flow again takes place. In particular, it follows from the
last formula of (\ref{4-31}) that one  is equivalent to
$R_*=\frac{\beta}{r}$  (here $\beta$ can be easily calculated) in a
vicinity of the point $(0;0)$. So,  the interpretation is the same
as for the exact solution (\ref{4-13**}).

\section{\bf Conclusions} \label{s6}


In this paper, the Lie symmetry analysis of the (1+2)-dimensional
nonlinear  BVP (\ref{2-1})--(\ref{2-2}), which is the
two-dimensional (in space) approximation  of the known   tumour
growth model proposed in \cite{by-ki-2003}, was carried out. The
symmetries derived are applied for the reduction of the nonlinear
BVPs  in question to those of lower dimensionality. Finally, the
reduced problems with  correctly-specified coefficients were exactly
solved and  the exact solutions derived were  analysed, in
particular, some plots were build in order to understand the
time-space behaviour of these solutions.

It should be noted that a {\it complete} Lie symmetry classification
of BVP (\ref{2-1})--(\ref{2-2}) is still an open problem because we
deal with
 a class of BVPs involving three arbitrary functions  $S, \  D$  and $\Sigma$,
 which can possess essentially  different forms.  Here this class
 was examined in details only in the power-law  case, which is
 the most common in such kind studies, and the general case, i.e.
 assuming that three above mentioned functions are arbitrary. We
 foresee  that Lie symmetry of the nonlinear  BVP (\ref{2-1})--(\ref{2-2})
 with correctly-specified functions $S, \  D$  and $\Sigma$ (not necessary
 of the form (\ref{2-5})!)  can be wider than one derived in
 Theorem~\ref{th-2-3}
  and are going to continue this research.

\section{\bf Acknowledgments}
The first author thanks  John R.King (University of Nottingham) for the fruitful discussions about the results presented in this paper and   the School of Mathematical sciences of the   University of Nottingham,
where  this work was initiated, for hospitality  and partial financial support.

\end{document}